\documentclass[12pt,draftcls,onecolumn]{IEEEtran}
\usepackage[utf8]{inputenc}

\usepackage{graphicx}
\usepackage{tikz} 
\usepackage{amsmath}
\usepackage{amssymb}
\usepackage{amsfonts}
\usepackage[version=4]{mhchem}
\usepackage{siunitx}
\usepackage{epstopdf}

\usepackage{longtable,tabularx}
\usepackage{accents}

\usetikzlibrary{arrows}
 
\ifCLASSINFOpdf
\else
\fi
\begin{document}
%
\title{Polynomial Chaos-Based Adaptive Control for Nonlinear
Systems}
%
%
%

\author{Mateus~de~Freitas~Virgilio~Pereira,
        Igor~Afonso~Acampora~Prado,
        Davi~Ferreira~de~Castro, and~Jos\'{e}~Manoel~Balthazar

\thanks{M.F.V. Pereira (mfvp@umich.edu), I.A.A. Prado (igorap@ita.br), D.F. de Castro (davifc@ita.br) and J.M. Balthazar (jmbaltha@ita.br) were with the Aeronautics Institute of Technology - ITA, Brazil.}
\thanks{Manuscript received xxxx}}

%
%

\markboth{}{}%
%



\maketitle

\begin{abstract}
In this work we present a nonlinear adaptive suboptimal control strategy for uncertain nonlinear systems. Stochastic parametric uncertainty is dealt with by employing spectral decomposition of the random variables by means of the generalized polynomial chaos expansion. The projection of uncertainty onto the orthogonal polynomial basis, prescribed by the Wiener-Askey scheme, provides a deterministic model from which the control laws are designed. We apply the nonlinear feedback control law to an automatic pilot system for recovering an aircraft with uncertain aerodynamic data from stall while providing acceptable dynamic response. The control law is obtained by approximating the Hamilton-Jacobi-Bellman equation by a pertubational procedure and asymptotic weak stability in probability of the controlled nonlinear system is verified in a deterministic way.
\end{abstract}


%
\IEEEpeerreviewmaketitle

\section{Introduction}

High-angle-of-attack flights have been a research topic of broad interest in aeronautical engineering for decades. Such flight regime is usual for modern high-performance aircraft, whose maneuverability and controllability should be possible even in the stall region. In this situation, the lift coefficient cannot be represented as a linear function of angle of attack and therefore nonlinear aerodynamic terms should be taken into consideration. In this context, the automatic pilot system should work toward recovering the aircraft from stall while ensuring appropriate dynamic response. Hence, the control system design should account for the nonlinear phenomena that affect the vehicle operation as well as the uncertainties in the vehicle model and environment. Therefore, in real implementations, such control system ought to be robust, that is, capable of working properly, within the designed specifications, even when subject to either disturbances or uncertainties in the determination of the vehicle parameters.

Adaptive and robust control are typical approaches for designing control structures for systems affected by uncertainties. Applications of such techniques are widespread in the aerospace engineering  literature. Recent works include nonlinear dynamic inversion with multivariate
spline-based adaptive control allocation to compensate for aerodynamic uncertainties \cite{Tol}, autopilot design using $H_{\infty}$ loop shaping for missiles at high angles of attack \cite{Mahmood}, adaptively augmented LQR controller for agile aircraft \cite{Zollistch}, $\mathcal{L}_1$ adaptive controller for a BWB aircraft \cite{Zheng}, $H_2 / H_{\infty}$ robust control for hypersonic vehicles \cite{Huang}, and Lypunov function-based adaptive control for longitudinal dynamics \cite{Gavilan}. Even though such techniques are well established in the area, they generally still present some unresolved issues. For instance, the mathematical proof of asymptotic stability in adaptive structures is usually fairly involved and, in some cases, not even possible. Robust control is too cautious for considering only the worst-case scenario for control design, thus being blind to the actual behavior of uncertainty in the system and possibly degrading performance and cost of the feedback scheme.  In view of these problems, a less conservative method for controlling nonlinear systems is sought, in a way that stability, optimality and robustness are all guaranteed \cite{Pereira}.
	
	The robustness of the controller to stochastic inputs can be assessed by conventional approaches such as Monte Carlo sampling methods, which might be highly computationally costly and not practical for a large number of random variables. Polynomial chaos expansion is an alternative to Monte Carlo methods, since it provides a simpler framework to deal with the propagation of uncertainty in the dynamical model. It is the spectral decomposition of a stochastic process in the random dimension in which it is parametrized. The random trial basis is the Askey-scheme based orthogonal polynomials. Introduced first by Wiener \cite{Wiener} for Gaussian random variables and then generalized by Xiu and Karniadakis \cite{Xiu2002} for other probability distributions, such decomposition is based on the Cameron and Martin Theorem \cite{Cameron}, which guarantees the convergence in $L_2$ sense for stochastic processes with finite second moment. Many researches have attested that spectral methods based on polynomial chaos expansions can be a computationally efficient alternative to expensive conventional approaches based on Monte Carlo sampling \cite{Xiu2002,Fisher2009,Prabhakar,Xiu2003}. The polynomial chaos expansion framework has been applied in sensitivity analysis \cite{Abdelkefi,Sudret}, uncertainty quantification \cite{Prabhakar,Sepahvand}, structural dynamics \cite{Ghanem,Jacquelin}, robust stability analysis \cite{Fisher20081,Nechak}, collision avoidance \cite{Kim}, fluid dynamics \cite{Hosder,Ko}, to cite a few.

	The use of polynomial chaos expansions in control theory is recent. Hover and Triantafyllou \cite{Hover} demonstrated that this method applies to the stability study of nonlinear systems, especially when the methods of Lyapunov are inconclusive and Monte Carlo simulations are expensive. Fisher and Bhattacharya \cite{Fisher2009,Fisher20081,Fisher20082} presented a generalized procedure for the stability analysis of linear and polynomial systems and a systematic framework for designing linear quadratic regulators (LQR) for stochastic linear systems using polynomial chaos expansions. Other applications in control theory are found in optimal control \cite{Peng,Sriram,Pereira,PereiraASME}, model predictive control \cite{Fagiano,Mesbah} and robust control \cite{Templeton}. The majority of the previous work only consider linear control theory, and the synergy between nonlinear techniques and polynomial chaos expansions has not been fully explored yet.
	
	In this paper, we are especially interested in the benefits that polynomial chaos expansions can bring to the design of a robust control strategy to recover the aircraft from the stall region while guaranteeing appropriate stability and performance. Many design challenges arise in flights at high incidence owing to nonlinear phenomena that might strike the aircraft performance. For instance, bifurcation behavior regarding the elevator control input may be observed, which may cause the jump phenomenon or lead the vehicle to instability \cite{Liaw,Hui,Jahnke}. Moreover, in an aerodynamic standpoint, the usual potential flow approach can no longer predict the forces and moments acting on the aircraft due to complicated phenomena such as boundary layer separation and vortex breakdown, to not mention compressibility \cite{Rom}. Incidentally, many aircraft models are unable to predict completely and satisfactorily the dynamics at high angles of attack. Therefore, for a reasonable model, nonlinearities and uncertainties present in the real operation of these vehicles have to be considered in order to avoid unstable regions of operation. Many control schemes for the automatic pilot system are found in the literature, such as nonlinear quadratic regulator \cite{Garrard2}, washout filters for bifurcation control \cite{LeeAbed}, dynamic inversion \cite{Bugajski}, probabilistic robust control \cite{Wang}, adaptive neural networks \cite{Calise}, to cite a few. Even though such control strategies presented good performance at controlling high incidence flights, they may have a difficult implementation and, in some cases, do not guarantee stability or robustness. The development of a nonlinear control strategy that incorporates polynomial chaos expansions to handle uncertainty holds the promise to overcome some of these issues.

This work has two fundamental objectives. First, to investigate the application of the polynomial chaos expansion framework in the design of control laws for nonlinear systems with uncertain stochastic parameters and, perforce, proofs of stability, robustness and optimality are sought. We also present an implementation strategy of the proposed controller that resembles adaptive control. The second objective is to design a controller that is capable of recovering the aircraft from the stall condition with adequate dynamic response even when the aerodynamic model is not fully known.  The foremost contribution of this paper lies in the novel formulation of nonlinear optimal control problems with stochastic parametric uncertainty in a deterministic framework whilst assuring stability and optimality, ergo addressing two of the most critical issues encountered in traditional approaches. Most importantly, the proposed method takes into consideration the a priori knowledge of the underlying probability of uncertainty during the control design, thus being less conservative than the conventional nonlinear robust control methods for systems with parametric uncertainty.  For assuming that the uncertainty affecting the control system has a stochastic nature and using such information for design purposes, the proposed control strategy can be classified under the generic name of probabilistic robust control \cite{Calafiore}. To the best knowledge of the author, this work is the first to present a general deterministic framework for designing feedback nonlinear optimal control laws for longitudinal attitude control of a nonlinear aircraft model with uncertain aerodynamic data by means of intrusive polynomial chaos expansions. 

This paper is organized as follows: Section \ref{Polynomial Chaos} reviews the polynomial chaos theory for propagating uncertainties through nonlinear systems. Section \ref{Control Design} derives the nonlinear optimal control strategy and the proofs of stability, and proposes an adaptive implementation scheme. The remaining sections delve into the aircraft modeling for longitudinal autopilot design. Section \ref{Aircraft Model} presents the equations of motion of the aircraft, as well as the aerodynamic data and the simulation model.  Chapter \ref{Simulations} presents the results obtained from numerical simulations of the controlled aircraft model. Finally, Chapter \ref{Conclusions} sums up the major conclusions from this work and suggests topics for further investigation.

\section{Polynomial Chaos Expansions}
\label{Polynomial Chaos}
In this section we introduce the polynomial chaos expansion for representing a stochastic process according to the Wiener's theory of homogeneous chaos \cite{Wiener}. We also describe the use of such framework to transform stochastic nonlinear dynamical systems into an augmented deterministic system for the decomposition coefficients.
\subsection{Spectral Decomposition and The Wiener-Askey Scheme}
\label{sec:1.1}
Define the set of multi-indices with finite number of nonzero components as
\begin{equation}
\mathcal{J}= \left\{ \pmb{\alpha} = (\alpha_{i})_{i\geq 1}, \alpha_{i} \in \mathbb{N},|\pmb{\alpha}|=\sum_{i=1}^{\infty} \alpha_{i}<\infty\right\}.
\end{equation}
For an index $\pmb{\alpha} \in \mathcal{J}$, the Wick polynomial of order $|\pmb{\alpha}|$ in an infinite number of independent and identically distributed normal random variables $\pmb{\xi}=(\xi_1,\xi_2,\ldots)$ is defined by
\begin{equation}
\Psi_{\pmb{\alpha}}(\pmb{\xi})=\prod_{i=0}^{\infty}H_{\alpha_{i}}(\xi_{i}),
\end{equation}
where $H_n(\cdot)$ is the normalized $n$-th order multivariate orthogonal Hermite polynomial. The random functions $\Psi_{\pmb{\alpha}}$ form a complete orthonormal basis in $L_2$ on the probability space with respect to the Gaussian measure generated by $\pmb{\xi}$ \cite{Branicki}. According to the Cameron-Martin theorem \cite{Cameron}, a finite-variance random variable $g(\tau,\pmb{\xi})$, where $\tau$ denotes any deterministic parameter (e.g. the time instant), has the following decomposition in Wick polynomials:
\begin{equation}
g(\tau,\pmb{\xi})=\sum_{\pmb{\alpha}\in \mathcal{J}}g_{\pmb{\alpha}}(\tau)\Psi_{\pmb{\alpha}}(\pmb{\xi}),
\label{PCE}
\end{equation}
where $g_{\pmb{\alpha}}(\tau)$ are the mode strengths given by
\begin{equation}
g_{\pmb{\alpha}}(\tau)=\frac{\langle g(\tau,\pmb{\xi}),\Psi_{\pmb{\alpha}}(\pmb{\xi})\rangle}{\langle \Psi_{\pmb{\alpha}}^2(\pmb{\xi})\rangle}.
\label{PCEgcoeff}
\end{equation}
Notation $\langle \cdot \rangle$ represents the multivariate inner product in $L_2 (D_\xi)$ space and $D_\xi$ is the domain of $\pmb{\xi}$. The spectral expansion in Eq. \ref{PCE} is referred to as the Polynomial Chaos Expansion (PCE) of the random variable $g$. Given a $g$ and a polynomial basis $\Psi_{\pmb{\alpha}}(\pmb{\xi})$, the expansion in Eq. \ref{PCE} is unique. A fundamental property of this expansion is that it converges in the $L_2$ sense. In practice, the PCE will be truncated to a finite number of terms by limiting the germ $\pmb{\xi}$ to $r$ normal random variables and the order of the multivariate polynomials $\Psi_{\pmb{\alpha}}$ to $|\pmb{\alpha}|\leq d$. The resulting number of terms in the truncated expansion is given by $P = \frac{(r+d)!}{r!d!}-1$. The parameter $P$ has to be chosen large enough for the approximation of Eq. \ref{PCE} to be accurate. The rate of convergence of the truncated PCE depends on the the smoothness of $g(\tau,\pmb{\xi})$ as a functional in a Hilbert measure space. Roughly speaking, for a PCE of order $P$, denoted by $g_P(\tau,\pmb{\xi})$, the approximation error $\Vert g(\tau,\pmb{\xi}) - g_P(\tau,\pmb{\xi}) \Vert$ is $\mathcal{O}(P^{-\mathfrak{p}})$, where $\mathfrak{p}$ is the differentiability of the function $g(\tau,\pmb{\xi})$. For an analytic function $g(\tau,\pmb{\xi})$, the convergence rate is exponential, i.e., $\Vert g(\tau,\pmb{\xi}) - g_P(\tau,\pmb{\xi})\Vert = \mathcal{O}(e^{-\varrho P})$ for some constant $\varrho>0$ \cite{Kim}.

	Xiu and Karniadakis \cite{Xiu2002} extended the PCE framework to non-Gaussian random inputs by elaborating a tree that maps a certain distribution to the corresponding orthogonal polynomial that guarantees the optimal convergence of the expansion. Table \ref{tab:1} shows the pairing for continuous and discrete distributed random variables. This broad framework is called Generalized Polynomial Chaos (gPC) expansion and it associates the probability density function of the germ $\pmb{\xi}$ with polynomials within the Askey-scheme that have a similar weight function. Such polynomials form a complete basis in the Hilbert space determined by their corresponding support \cite{Fisher2009,Askey}.  For arbitrary uncertain input distributions outside the Askey scheme, the Gram-Schmidt orthogonalization algorithm can be used to generate a polynomial basis to achieve exponential convergence of the expansion \cite{Witteveen}. Note that, in the gPC framework, the random variables in $\pmb{\xi}$ do not need to have the same probability distribution function, thus we can propagate simultaneously through the model uncertainties with different distributions, as long as we built the appropriate multivariate tensor product polynomial for the decomposition basis.

\begin{table}[!h]
\centering
\caption{The Wiener-Askey scheme.}
\label{tab:1} 
\begin{tabular}{lll}
\hline\noalign{\smallskip}
Random variable $\xi$ & Polynomial basis $\Psi_{k}$ & Support  \\
\noalign{\smallskip}\hline\noalign{\smallskip}
Gaussian & Hermite & $(-\infty, \infty)$\\
Uniform & Legendre & $[a,b]$\\
Gamma & Laguerre & $[0,\infty)$\\
Beta & Jacobi & $[a,b]$ \\
\noalign{\smallskip}\hline
\end{tabular}
\end{table}

	The gPC method is especially useful in solving or analyzing stochastic differential equations. In control theory, the use of conventional methods to study the stability and to design a nonlinear robust controller subject to random inputs can lead, in general, to inconclusive results or does not cover all the possible scenarios in an efficient way. In this sense, the gPC is a powerful tool since the stability characteristic of a dynamical system can be inferred from the decay of the modes strengths over time \cite{Fisher20081,Hover}, that is, in a deterministic framework. 
	
	We should stress here that, in this work, we only consider the application of gPC expansions to deal with stochastic systems modeled by parametric (or predictable) stochastic differential equations, i.e. stochastic processes that are fully specified in terms of the random variables $\pmb{\xi}$. Such stochastic processes are completely determined for $t>t_0$ if the past, $t\leq t_0$, is known \cite{Papoulis}. That is the case for dynamical systems with stochastic parametric uncertainty, such as the aircraft model presented in Section \ref{Aircraft Model}.
	
\subsection{Expansion of Stochastic Differential Nonlinear Equations}
\label{sec:1.2}
Consider stochastic nonlinear dynamical systems of the following form:
\begin{equation}
\pmb{\dot{x}}(t,\pmb{\xi}) = \pmb{A}(\pmb{\xi})\pmb{x}(t,\pmb{\xi})+\pmb{h}(\pmb{x},\pmb{\xi})+\pmb{B}(\pmb{\xi})\pmb{u}(t,\pmb{\xi}),
\label{StochSys}
\end{equation}
where $\pmb{x} \in \mathbf{R}^{n}$ is the state vector, $\pmb{u} \in \mathbf{R}^{m}$ is the control input, $\pmb{A} \in \mathbf{R}^{n \times n}$ and $\pmb{B} \in \mathbf{R}^{n \times m}$ are matrices and $\pmb{h} \in \mathbf{R}^{n}$ is a vector whose components are continuous nonlinear functions of $\pmb{x}$. The random variable $\pmb{\xi}=(\xi_1,\xi_2,\ldots,\xi_r )$ represents uncertainties, with a known stationary distribution, by which the parameters of the model or the initial conditions are expressed. 

	Denote the components of $\pmb{x}(t,\pmb{\xi})$, $\pmb{u}(t,\pmb{\xi})$ and $\pmb{h}(\pmb{x},\pmb{\xi})$ by, respectively, $x_i (t,\pmb{\xi})$, $u_i (t,\pmb{\xi})$ and $h_i (\pmb{x},\pmb{\xi})$. Also, denote the elements of $\pmb{A}(\pmb{\xi})$ and $\pmb{B}(\pmb{\xi})$ by, respectively, $A_{ij}(\pmb{\xi})$ and $B_{ij}(\pmb{\xi})$.  We can represent $x_i (t,\pmb{\xi})$, $u_i (t,\pmb{\xi})$, $A_{ij} (\pmb{\xi})$, $B_{ij}(\pmb{\xi})$ and $h_i (\pmb{x},\pmb{\xi})$ by means of the gPC expansion of order $P$ in the orthogonal polynomial basis $\Psi_{k}(\pmb{\xi})$ according to the Wiener-Askey scheme (see Table \ref{tab:1}):
\begin{equation}
x_i(t,\pmb{\xi})\approx \sum_{k=0}^{P} x_{i,k}(t)\Psi_{k}(\pmb{\xi}), 
\label{gPCx}
\end{equation}	
\begin{equation}
u_i(t,\pmb{\xi})\approx \sum_{k=0}^{P} u_{i,k}(t)\Psi_{k}(\pmb{\xi}), 
\label{gPCu}
\end{equation}	
\begin{equation}
A_{ij}(\pmb{\xi})\approx \sum_{k=0}^{P} a_{ij,k}\Psi_{k}(\pmb{\xi}), 
\end{equation}	
\begin{equation}
B_{ij}(\pmb{\xi})\approx \sum_{k=0}^{P} b_{ij,k}\Psi_{k}(\pmb{\xi}),
\end{equation}
\begin{equation}
h_i(\pmb{x},\pmb{\xi})\approx \sum_{k=0}^{P} h_{i,k}(t)\Psi_{k}(\pmb{\xi}). 
\label{gPCh}
\end{equation}	
Note that, as long as the relation between $\pmb{h}$ and $\pmb{x}$ is known, the expansion for $h_i (\pmb{x},\pmb{\xi})$, Eq. \ref{gPCh}, may be reformulated in terms of the expansions $x_i (t,\pmb{\xi})$. 

Let us define the following vector notation for the mode strengths appearing in Eq. \ref{gPCx}-\ref{gPCh}:
\begin{equation}
\pmb{x}_i \triangleq \left[ \begin{array}{cccc} x_{i,0}(t) & x_{i,1}(t) & \ldots & x_{i,P}(t) \end{array} \right]^T \in \mathbb{R}^{P+1},
\end{equation}
\begin{equation}
\pmb{u}_i \triangleq \left[ \begin{array}{cccc} u_{i,0}(t) & u_{i,1}(t) & \ldots & u_{i,P}(t) \end{array} \right]^T \in \mathbb{R}^{P+1},
\end{equation}
\begin{equation}
\pmb{h}_i \triangleq \left[ \begin{array}{cccc} h_{i,0}(t) & h_{i,1}(t) & \ldots & h_{i,P}(t) \end{array} \right]^T \in \mathbb{R}^{P+1}.
\label{vectorh}
\end{equation}
Define also the matrices $\pmb{A}_k \in \mathbb{R}^{(n\times n)} \mbox{ with } \{A_k\}_{ij} = a_{ij,k}$ and $\pmb{B}_k \in \mathbb{R}^{(n\times m)} \mbox{ with } \{B_k\}_{ij} = b_{ij,k}$ for $k=0,\ldots,P$.
Using the intrusive approach in \cite{Xiu2002}, the coefficients $a_{ij,k}$ and $b_{ij,k}$ are computed by the Galerkin projection of each mode onto the polynomial basis $\{\Psi_q \}_{q=0}^{P}$, in order to ensure the error is orthogonal to the functional space spanned by the finite-dimensional basis $\Psi_k$:
\begin{equation}
a_{ij,k} = \frac{\langle A_{ij}(\pmb{\xi}),\Psi_{k}(\pmb{\xi}) \rangle}{\langle \Psi_{k}^{2}(\pmb{\xi})\rangle},
\end{equation}
\begin{equation}
b_{ij,k} = \frac{\langle B_{ij}(\pmb{\xi}),\Psi_{k}(\pmb{\xi}) \rangle}{\langle \Psi_{k}^{2}(\pmb{\xi})\rangle}.
\end{equation}
The modes strengths $\pmb{x}_i$, $\pmb{u}_i$ and $\pmb{h}_i$ are also worked out in an intrusive way by employing the Galerkin projection. The gPC expansions from Eq. \ref{gPCx} to \ref{gPCh} are substituted into the original system, Eq. \ref{StochSys}, and the projection with respect to the polynomial basis is taken, yielding a system of $n(P+1)$ deterministic ordinary differential equations for the modes, which, as shown by Fisher and Bhattacharya \cite{Fisher2009} and Pereira et al. \cite{Pereira}, can be put into the following suitable form:
\begin{equation}
\pmb{\dot{\mathcal{X}}} = \pmb{\mathcal{A}}\pmb{\mathcal{X}}+\pmb{\mathcal{H}}+\pmb{\mathcal{B}}\pmb{\mathcal{U}},
\label{FinalPC}
\end{equation}

where $\pmb{\mathcal{X}} \in \mathbb{R}^{n(P+1)}$, $\pmb{\mathcal{H}} \in \mathbb{R}^{n(P+1)}$ and $\pmb{\mathcal{U}} \in \mathbb{R}^{m(P+1)}$ are given by
\begin{equation}
\pmb{\mathcal{X}} = \left[ \begin{array}{cccc} \pmb{x}_{1}^T & \pmb{x}_{2}^T & \ldots &\pmb{x}_{n}^T\end{array} \right]^T,
\end{equation}
\begin{equation}
\pmb{\mathcal{H}} = \left[ \begin{array}{cccc} \pmb{h}_{1}^T & \pmb{h}_{2}^T & \ldots &\pmb{h}_{n}^T\end{array} \right]^T.
\label{bigH}
\end{equation}
\begin{equation}
\pmb{\mathcal{U}} = \left[ \begin{array}{cccc} \pmb{u}_{1}^T & \pmb{u}_{2}^T & \ldots &\pmb{u}_{m}^T\end{array} \right]^T.
\end{equation}

Matrices $\pmb{\mathcal{A}} \in \mathbb{R}^{n(P+1)\times n(P+1)}$ and $\pmb{\mathcal{B}} \in \mathbb{R}^{n(P+1)\times m(P+1)}$ are defined as:
\begin{equation}
\pmb{\mathcal{A}} = \sum_{k=0}^{P} \pmb{A}_k \otimes \pmb{E}_k,
\end{equation}
\begin{equation}
\pmb{\mathcal{B}} = \sum_{k=0}^{P} \pmb{B}_k \otimes \pmb{E}_k,
\end{equation}
where $\pmb{A}_k$ and $\pmb{B}_k$ are defined below Eq. \ref{vectorh}, $\otimes$ denotes the Kronecker product and  $\pmb{E}_k \in \mathbb{R}^{(P+1)\times(P+1)}$ is a symmetric matrix given by:
\begin{equation}
\pmb{E}_k = \left[ \begin{array}{cccc} \hat{e}_{1k1} & \hat{e}_{1k2} & \ldots \hat{e}_{1kP}\\ \hat{e}_{1k2} & \hat{e}_{2k2} & \ldots \hat{e}_{2kP} \\ \vdots & \vdots & \ddots & \vdots \\ \hat{e}_{1kP} & \hat{e}_{2kP} & \ldots \hat{e}_{PkP}  \end{array} \right].
\end{equation}
The nonlinearities in vector $\pmb{h}(\pmb{x},\pmb{\xi})$ are firstly expanded in Taylor series around the mean of the argument in order to obtain an approximate polynomial form, such as in
\begin{equation}
h_{i}(x_j(\pmb{\xi})) \approx \sum_{p=0}^{N} \frac{h_{i}^{(p)}(x_{j,0})}{p!}\left( \sum_{k=1}^{P} x_{j,k} \Psi_k (\pmb{\xi}) \right)^{p},
\label{HnonPoly}
\end{equation}
and then the Galerkin projection is taken in order to put the augmented vector in form of Eq. \ref{bigH}. For nonsmooth nonlinearities, non-intrusive approaches based on sampling can be used, as discussed by Dubusschere et al. \cite{Debusschere}, and will not be covered in this work.

\section{Control Design}
\label{Control Design}
In this section we present the incorporation of the polynomial chaos framework for designing control laws for nonlinear systems with stochastic parametric uncertainty. Section \ref{Stability} presents some definitions about stability of stochastic dynamical systems. In Sections \ref{Optimal Control Design} and \ref{Nonlinear Suboptimal Control}, we introduce a suboptimal control strategy for regulation of uncertain nonlinear systems using polynomial chaos expansions. In Section \ref{Implementation}, an implementation scheme is proposed.
\subsection{Stability of Stochastic Dynamical Systems}
\label{Stability}

Here we extend to nonlinear systems the polynomial chaos-based framework introduced by Fisher and Bhattacharya \cite{Fisher2009} for linear systems to analyze the stability of dynamical systems with stochastic parameters. Consider a general nonlinear stochastic system with stochastic parametric uncertainty:
\begin{equation}
\dot{\pmb{x}}(t,\pmb{\xi}) = \pmb{f}(t,\pmb{x},\pmb{\xi}) ~~~~~~~~~~~~~~ \pmb{x}(t_0,\pmb{\xi}) = \pmb{x}_0
\label{StochDef}
\end{equation}
where $\pmb{x} \in \mathbb{R}^n$ is the state vector and $\pmb{\xi} \in \mathbb{R}^d$ is the vector of random variables. Following the usual procedure of introducing new variables, equal to the deviations of the corresponding coordinates of the perturbed motion from their unperturbed values \cite{Khasminskii}, only the stability of the solution $\pmb{x}(t,\pmb{\xi}) \equiv 0$ has to be considered as long as the following condition is satisfied:
\begin{equation}
\pmb{f}(t,\pmb{0},\pmb{\xi}) = \pmb{0} ~~~~~~~~~~~~~~~~~ \mbox{for all }t>0.
\end{equation}

Consider the following definitions of stochastic stability of the zero equilibrium of the system in Eq. \ref{StochDef}  \cite{ChenChen,Kushner,Khasminskii}:
\paragraph*{Definition} Stability in probability: The solution $\pmb{x}(t,\pmb{\xi})$ is said to be stable in probability  for $t \geq t_0$ if for any $\epsilon>0$
\begin{equation}
\lim_{\pmb{x}_0 \to \pmb{0}}\rm{Pr} \left\{\sup_{t \geq t_0}|\pmb{x(t)}| \geq \epsilon \right\}=0.
\label{strong}
\end{equation}
In addition, the solution is said to be weakly stable in probability if
\begin{equation}
\lim_{\pmb{x}_0 \to \pmb{0}} \sup_{t\geq t_0} \rm{Pr} \left\{|\pmb{x(t)}| \geq \epsilon \right\}=0 ~~~~~~~~ \mbox{for all } \epsilon>0.
\label{weak}
\end{equation}
In the particular case of linear stochastic differential equations, the two definitions are equivalent \cite{Khasminskii}.

We should also consider the definition of $p$-stability, i.e., the stability of the moments of the stochastic process $\{ \pmb{x}(t) \}$. It consists of the study of growth or decay of the moments of the solution of Eq. \ref{StochDef}, which are deterministic functions \cite{ChenChen,Khasminskii}.
\paragraph*{Definition} $p$-stability: The solution $\pmb{x}(t,\pmb{\xi})$ is said to be $p$-stable for $p>0$ and $t\geq t_0$ if, for every $\epsilon>0$, there exists a $\delta>0$ such that
\begin{equation}
\sup_{t\geq t_0} E[|\pmb{x}(t)|^p]\leq \epsilon, ~~~~~~~~ \mbox{for all } \pmb{x}_0 : |\pmb{x}_0|\leq \delta.
\end{equation}
It is said to be asymptotically $p$-stable if it is $p$-stable and moreover
\begin{equation}
\lim_{t \to \infty} E[|\pmb{x}(t)|^p]=0, ~~~~~~~~ \mbox{for all } \pmb{x}_0 : |\pmb{x}_0|\leq \delta.
\end{equation}

If the last definition holds for any $\pmb{x}_0$ in the domain of $\pmb{f}(t,\pmb{x},\pmb{\xi})$, then we say that the origin is asymptotically stable in the large. Furthermore, if $\delta$ does not depend on the $t_0$, then the origin is said to be uniformly asymptotically stable. Now consider the truncated polynomial chaos expansion of the solution $\pmb{x}(t)$ of the nonlinear stochastic differential equation in Eq. \ref{StochDef}. For the sake of simplicity and with no loss of generality, we consider the one-dimensional case, i.e., $n=1$ and $d=1$:
\begin{equation}
x(t,\xi) =\pmb{\mathcal{X}}^T \pmb{\Psi}
\label{xuni}
\end{equation}
such that
\begin{equation}
\dot{\pmb{\mathcal{X}}}(t) = f\left(t,\pmb{\mathcal{X}} \right) ~~~~~~~~~~~~~~ \pmb{\mathcal{X}}(t_0) = \pmb{\mathcal{X}}_0,
\label{GeneralPCexp}
\end{equation}
where $\pmb{\mathcal{X}} = \left[ \begin{array}{cccc} x_{0}(t) & x_{1}(t) & \ldots & x_{P}(t) \end{array} \right]^T \in \mathbb{R}^{P+1}$ is the vector that concatenates the gPC mode strengths, and $\pmb{\Psi} = \left[ \begin{array}{cccc} \Psi_{0}(\xi) & \Psi_{1}(\xi) & \ldots & \Psi_{P}(\xi) \end{array} \right]^T \in \mathbb{R}^{P+1}$ is the vector with the polynomial bases. The following theorem presents, using the polynomial chaos framework, the conditions for weak stochastic stability of the zero equilibrium of Eq. \ref{StochDef}.
\paragraph*{Theorem 3.1} The zero equilibrium solution of the nonlinear dynamical system with stochastic parametric uncertainty in Eq. \ref{StochDef} is uniformly weakly asymptotically stable in probability in the large if, for the augmented deterministic system in Eq. \ref{GeneralPCexp} for the mode strengths of the polynomial chaos expansion of Eq. \ref{StochDef}, a scalar function $\mathcal{V}(t,\pmb{\mathcal{X}})$ with continuous first partial derivatives with respect to $\pmb{\mathcal{X}}$ and $t$ exists such that:

(i) $\mathcal{V}(t,\pmb{0}) = 0$;

(ii) $\mathcal{V}(t,\pmb{\mathcal{X}})$ is positive-definite, i.e., there exists a continuous nondecreasing scalar function $\kappa_1(|\pmb{\mathcal{X}}|)$ such that $\kappa_1(\pmb{0})=0$ and $\mathcal{V}(t,\pmb{\mathcal{X}}) \geq \kappa_1(|\pmb{\mathcal{X}}|) > 0$ for all $t$ and all $\pmb{\mathcal{X}} \neq \pmb{0}$;

(iii) $\mathcal{V}(t,\pmb{\mathcal{X}})$ is decrescent, i.e., there exists a continuous nondecreasing scalar function $\kappa_2(|\pmb{\mathcal{X}}|)$ such that $\kappa_2(\pmb{0})=0$ and $\kappa_2(|\pmb{\mathcal{X}}|) \geq \mathcal{V}(t,\pmb{\mathcal{X}})$ for all $t$;

(iv) $\dot{\mathcal{V}}(t,\pmb{\mathcal{X}})$ is negative definite, that is,
\begin{equation*}
\dot{\mathcal{V}}(t,\pmb{\mathcal{X}}) = \frac{\partial \mathcal{V}}{\partial t} + (\bigtriangledown \mathcal{V})^T f(t,\pmb{\mathcal{X}}) \leq -\kappa_3(|\pmb{\mathcal{X}}|)<0,
\end{equation*}
where $\kappa_3(|\pmb{\mathcal{X}}|)$ is a continuous nondecreasing scalar function such that $\kappa_3(\pmb{0})=0$;

(v) $\mathcal{V}(t,\pmb{\mathcal{X}})$ is radially unbounded, i.e., $\kappa_1(|\pmb{\mathcal{X}}|) \to \infty$ with $|\pmb{\mathcal{X}}| \to \infty$ .

\noindent
\emph{Proof:} We split the proof into two parts:

(a) Stability of the augmented deterministic system: First note that the selection of the Lyapunov function that satisfies conditions (i)-(v) provides sufficient conditions for the uniformly asymptotically stability in the large of the solution $\pmb{\mathcal{X}} \equiv 0$ of Eq. \ref{GeneralPCexp}, that is, for all $t\geq t_0$ and all $\pmb{\mathcal{X}}_0 \in \mathbb{R}^{P+1}$ there is $\delta>0$ such that, if $|\pmb{\mathcal{X}}(t_0)|<\delta$, then $|\pmb{\mathcal{X}}(t)|\to 0$ as $t \to \infty$, and the convergence is uniform with respect to $t_0$, i.e., for any $\sigma>0$ there exists $g(\sigma)<\infty$ such that $|\pmb{\mathcal{X}}(t_0)|<\delta$ implies $|\pmb{\mathcal{X}}(t+t_0)|<\sigma$ for all $t>g(\sigma)$. 

(b)$p$-stability of the stochastic system: This part uses part of the derivations presented by Fisher and Bhattacharya \cite{Fisher2009} in the analysis of $p$-stability. Recall that the vector $\pmb{\mathcal{X}}(t)$ contains the gPC mode strengths of the random variable $x(t,\xi)$ in a time $t$, where $\xi$ is a random variable with known continuous probability density distribution $f(\xi)$ over the support $I$. In face of Eq. \ref{xuni}, the $p$-th moment of $x(t,\xi)$ is given by:
\begin{equation}
\mu_p (t) = \sum_{i_1 =0}^{P} \cdots \sum_{i_p =0}^{P}x_{i_1}(t) \cdots x_{i_p}(t) \int_{I}\Psi_{i_1}(\xi) \cdots \Psi_{i_p}(\xi)f(\xi)d\xi,
\label{allmoments}
\end{equation}
for $p>0$. Note that the integral in Eq. \ref{allmoments} is finite for any $p$ and any set of orthogonal polynomials. Therefore, if conditions (i)-(v) are true, then $\pmb{\mathcal{X}}(t)$ is bounded, which implies that $x_i(t)$ is also bounded and consequently, all moments are bounded. Hence, the zero equilibrium of Eq. \ref{StochDef} is $p$-stable for all $p>0$ and $t\geq 0$. Moreover, if conditions (i)-(v) are true, then the solution $\pmb{\mathcal{X}}(t) \equiv \pmb{0}$ is asymptotically stable and $\pmb{\mathcal{X}}(t) \to 0$ as $t \to \infty$, that is, $x_i(t) \to 0$ and consequently $\mu_p(t) \to 0$ as $t \to \infty$ for all $p>0$ and $t\geq 0$. Finally, since the convergence of $\pmb{\mathcal{X}}$ holds for all $t_0$ and all $\pmb{\mathcal{X}}_0 \in \mathbb{R}^{P+1}$, then the convergence of $x_i(t)$ to the origin  holds for all $t_0$ and all $x_i(t_0) \in \mathbb{R}$ and thus the convergence of $\mu_p(t)$ is uniform and the system is $p$-stable in the large for all $p>0$ and $t\geq 0$. In sum, the zero equilibrium of Eq. \ref{StochDef} is uniformly asymptotically $p$-stable in the large for all $p>0$. Using the Chebyshev Inequality 
\begin{equation}
\sup_{t\geq t_0} \rm{Pr}\left\{ |\pmb{x}|(t) \geq \epsilon\right\} \leq \frac{1}{\epsilon^p} \sup_{t \geq t_0} E[|\pmb{x}(t)|^p]
\label{Chebyshev}
\end{equation}
and the squeezing principle, we have that the zero equilibrium of the nonlinear dynamical system with stochastic parametric uncertainty in Eq. \ref{StochDef} is uniformly weakly asymptotically stable in probability in the large. $\blacksquare$ 

In light of the Theorem 3.1, the goal of the control design is to find a control law that will drive the gPC mode strengths of Eq. \ref{StochDef} to the origin, thus guaranteeing the weak stochastic stability of the equilibria of the nonlinear dynamical system with stochastic parametric uncertainties.

\subsection{Optimal Control Design}
\label{Optimal Control Design}

The objective is to design a feedback control system that drives the trajectories of Eq. \ref{StochSys} from a certain initial condition to a small neighborhood around the origin, i.e. $\pmb{x}(t_f )=\pmb{0}$, while minimizing a performance index. 
We formulate the optimal control problem as follows:

\emph{Optimal Control Problem:} Given the nonlinear dynamics in Eq. \ref{StochSys}, with initial condition $\pmb{x}(t_0,\pmb{\xi})=\pmb{x}_0$, find the feedback control law $\pmb{u}(\pmb{x})$ that minimizes the cost function:
\begin{equation}
\min_{\pmb{u}}J(\pmb{x,u}) = \min_{\pmb{u}}E\left[ \int_{t_0}^{t_f} (l(\pmb{x})+\pmb{u}^T \pmb{R} \pmb{u})dt\right],
\label{cost}
\end{equation}
where
\begin{equation}
l(\pmb{x}) = \pmb{x}^T \pmb{Q} \pmb{x} +\epsilon N(\pmb{x}),
\end{equation}
$\pmb{Q} \in \mathbb{R}^{n\times n}$ is a real symmetric and positive definite matrix and  $\pmb{R} \in \mathbb{R}^{m\times m}$ is a real positive definite matrix. $N(\pmb{x})$ is a term that will be chosen later to account for the nonlinearities of the model and $\epsilon$ is a dimensionless parameter introduced for notation purposes. Using the generalized polynomial chaos expansion, the stochastic dynamical system can be represented as in Eq. \ref{FinalPC}:
\begin{equation}
\pmb{\dot{\mathcal{X}}} = \pmb{\mathcal{A}}\pmb{\mathcal{X}}+\epsilon \pmb{\mathcal{H}}+\pmb{\mathcal{B}}\pmb{\mathcal{U}}
\label{FinalPC2}
\end{equation}
Likewise, note that we can write the functional in Eq. \ref{cost} in the gPC framework, just as presented in Section \ref{Polynomial Chaos}. Using the orthogonal polynomials property, one can show that \cite{Fisher2009}:
\begin{equation}
E[\pmb{x}^T \pmb{Q} \pmb{x}] = \pmb{\mathcal{X}}^T \pmb{\widetilde{Q}}\pmb{\mathcal{X}},
\end{equation}
\begin{equation}
E[\pmb{u}^T \pmb{R} \pmb{u}] = \pmb{\mathcal{U}}^T \pmb{\widetilde{R}}\pmb{\mathcal{U}},
\end{equation}
where $\pmb{\widetilde{Q}}\triangleq \pmb{Q} \otimes \pmb{W}$, $\pmb{\widetilde{R}}\triangleq \pmb{R} \otimes \pmb{W}$ and $\pmb{W} \in \mathbb{R}^{(P+1)\times (P+1)}$ is a matrix whose elements are $\{\pmb{W}\}_{ij}=\langle \Psi_i(\pmb{\xi}),\Psi_j(\pmb{\xi}) \rangle$. In general, the expectation of $N(\pmb{x})$ can be written as a function of the gPC coefficients for $\pmb{x}$:
\begin{equation}
E[N(\pmb{x})] = \mathcal{N}(\pmb{\mathcal{X}}).
\end{equation}
Hence, the cost function in Eq. \ref{cost} can be represented in a deterministic form as:
\begin{equation}
\min_{\pmb{\mathcal{U}}}\widetilde{J}\left(  \pmb{\mathcal{X}},\pmb{\mathcal{U}} \right) = \min_{\pmb{\mathcal{U}}} \int_{t_0}^{t_f} \left( \mathcal{L}+\pmb{\mathcal{U}}^T \pmb{\widetilde{R}} \pmb{\mathcal{U}} \right) dt,
\label{cost2}
\end{equation}
where
\begin{equation}
\mathcal{L} = \pmb{\mathcal{X}}^T \pmb{\widetilde{Q}} \pmb{\mathcal{X}}+ \epsilon \mathcal{N}(\pmb{\mathcal{X}}).
\label{L11}
\end{equation}

Here we impose that $\pmb{\mathcal{H}}=\pmb{0}$ for $\pmb{\mathcal{X}}=\pmb{0}$, so that the origin is an equilibrium point. Otherwise a change of variables should be made in order to satisfy this condition. Note that solving the optimal control problem in Eq. \ref{cost} is equivalent to solving the problem in Eq. \ref{cost2}. Moreover, as discussed in Section \ref{Stability}, the stability of the augmented system guarantees that the original system is also stable in probability.

Let $V( \pmb{\mathcal{X}}) = \min \widetilde{J}\left(  \pmb{\mathcal{X}},\pmb{\mathcal{U}} \right)$ be the cost-to-go function, which is assumed to be continuously differentiable in $\pmb{\mathcal{X}}$, then the Hamilton-Jacobi-Bellman (HJB) equation provides sufficient conditions for optimality \cite{Bryson}: 
\begin{equation}
\begin{split}
\min_{\pmb{\mathcal{U}}}\left[ \frac{\partial V(\pmb{\mathcal{X}})}{\partial t}+ \mathcal{L}+\pmb{\mathcal{U}}^T \pmb{\widetilde{R}}\pmb{\mathcal{U}}+ \left( \frac{\partial V(\pmb{\mathcal{X}})}{\partial\pmb{\mathcal{X}}} \right)^T \left( \pmb{\mathcal{A}}\pmb{\mathcal{X}}+\epsilon \pmb{\mathcal{H}}+\pmb{\mathcal{B}}\pmb{\mathcal{U}} \right)\right] = 0.
\end{split}
\label{HJB}
\end{equation}
The general solution of this optimal control problem is of the form \cite{Garrard1967}:
\begin{equation}
\pmb{\mathcal{U}} = f\left( \pmb{\mathcal{X}},  \frac{\partial V(\pmb{\mathcal{X}})}{\partial\pmb{\mathcal{X}}} \right).
\end{equation}
Thus, it is necessary to solve a first order nonlinear partial differential equation.

\subsection{Nonlinear Suboptimal Control}
\label{Nonlinear Suboptimal Control}

Here we follow the formulation presented by  Garrard et al. \cite{Garrard1967} to design a nonlinear suboptimal control to drive the states of the augmented system to the origin in an infinite time horizon. From the HBJ equation (Eq. \ref{HJB}), it is straightforward to see that the optimal control law that minimizes the cost function in Eq. \ref{cost2} is of form:
\begin{equation}
\pmb{\mathcal{U}} = -\frac{1}{2}\pmb{\widetilde{R}}^{-1}\pmb{\mathcal{B}}^T \frac{\partial V(\pmb{\mathcal{X}})}{\partial\pmb{\mathcal{X}}}
\label{ControlNonEx}
\end{equation}
The substitution of the control law in Eq. \ref{ControlNonEx} into Eq. \ref{HJB} yields:
\begin{equation}
\begin{split}
\frac{\partial V}{\partial t} + \frac{\partial V}{\partial\pmb{\mathcal{X}}}^T \left( \pmb{\mathcal{A}}\pmb{\mathcal{X}}+\epsilon \pmb{\mathcal{H}} \right) - \frac{1}{4}   \frac{\partial V}{\partial\pmb{\mathcal{X}}} ^T  \pmb{\mathcal{B}}\pmb{\widetilde{R}}^{-1}\pmb{\mathcal{B}}^T  \frac{\partial V}{\partial\pmb{\mathcal{X}}}  \\+ \pmb{\mathcal{X}}^T \pmb{\widetilde{Q}} \pmb{\mathcal{X}}+ \epsilon \mathcal{N}(\pmb{\mathcal{X}}) =0
\end{split}
\label{HJB2}
\end{equation}
An analytical solution for the HJB equation in Eq. \ref{HJB2}, in general, does not exist. However, it is possible to obtain a control close to the optimal if it approximately satisfies the conditions specified in the optimal control problem. In a practical standpoint, a suboptimal solution can be chosen so that it satisfies criteria such as ease of implementation and reliability. Garrard et al. \cite{Garrard1967,Garrard} proposes a pertubational procedure to obtain the approximate solution of the optimal control problem. Assume a formal power series expansion of the cost-to-go function:
\begin{equation}
V(\pmb{\mathcal{X}}) = \sum_{i=2}^{\infty} \epsilon^{i-2} V_{i}(\pmb{\mathcal{X}})
\label{PowerSeries}
\end{equation}
Substituting Eq. \ref{PowerSeries} into Eq. \ref{HJB2} and equating the powers of $\epsilon$ to zero, we obtain \cite{Garrard1967}:
\begin{equation}
\begin{split}
\frac{\partial V_2}{\partial t}+\frac{\partial V_2}{\partial \pmb{\mathcal{X}}}^T\pmb{\mathcal{A}}\pmb{\mathcal{X}} -\frac{1}{4} \frac{\partial V_2}{\partial \pmb{\mathcal{X}}}^T \pmb{\mathcal{B}} \pmb{\widetilde{R}}^{-1} \pmb{\mathcal{B}}^T  \frac{\partial V_2}{\partial \pmb{\mathcal{X}}}\\ + \pmb{\mathcal{X}}^T \pmb{\widetilde{Q}} \pmb{\mathcal{X}}=0,
\end{split}
\end{equation}
\begin{equation}
\begin{split}
\frac{\partial V_3}{\partial t}+\frac{\partial V_3}{\partial \pmb{\mathcal{X}}}^T\pmb{\mathcal{A}}\pmb{\mathcal{X}} -\frac{1}{4} \frac{\partial V_2}{\partial \pmb{\mathcal{X}}}^T \pmb{\mathcal{B}} \pmb{\widetilde{R}}^{-1} \pmb{\mathcal{B}}^T  \frac{\partial V_3}{\partial \pmb{\mathcal{X}}}\\ -\frac{1}{4} \frac{\partial V_3}{\partial \pmb{\mathcal{X}}}^T \pmb{\mathcal{B}} \pmb{\widetilde{R}}^{-1} \pmb{\mathcal{B}}^T  \frac{\partial V_3}{\partial \pmb{\mathcal{X}}} + \mathcal{N}(\pmb{\mathcal{X}})=0,
\end{split}
\end{equation}
\begin{equation}
\begin{split}
\frac{\partial V_i}{\partial t}+\frac{\partial V_i}{\partial \pmb{\mathcal{X}}}^T\pmb{\mathcal{A}}\pmb{\mathcal{X}}+ \frac{\partial V_{i-1}}{\partial \pmb{\mathcal{X}}}^T \pmb{\mathcal{H}}\\ -\frac{1}{4} \sum^{n+2}_{ k\geq 2 ,~ l \geq 2} \frac{\partial V_k}{\partial \pmb{\mathcal{X}}}^T \pmb{\mathcal{B}} \pmb{\widetilde{R}}^{-1} \pmb{\mathcal{B}}^T  \frac{\partial V_l}{\partial \pmb{\mathcal{X}}} =0,
\end{split}
\label{labGen}
\end{equation}
where in the summation in Eq. \ref{labGen}, we require $k+l = i+2$. In order to determine $V_i$, these equation have to be solved successively. A special case occurs if, in Eq. \ref{L11}, we let $\mathcal{N}(\pmb{\mathcal{X}}) = 0$, so that $\mathcal{L}$ is a quadratic function in $\pmb{\mathcal{X}}$:
\begin{equation}
\mathcal{L} = \pmb{\mathcal{X}}^T \pmb{\widetilde{Q}} \pmb{\mathcal{X}}, 
\label{eqG}
\end{equation}
where  $\pmb{\widetilde{Q}} \in \mathbb{R}^{n(P+1) \times n(P+1)}$ is a real symmetric positive definite matrix. Assuming also that $\pmb{\mathcal{H}}$ has a polynomial form in $\pmb{\mathcal{X}}$, such that
\begin{equation}
\pmb{\mathcal{H}} = \sum_{\mathfrak{n}=1}^{N} \pmb{f}_{\mathfrak{n}+1}(\pmb{\mathcal{X}})
\label{polynomialH}
\end{equation}
where $f_{\mathfrak{n}+1}$ is of order $\mathfrak{n}+1$ in $\pmb{\mathcal{X}}$. Then the $V_{\mathfrak{n}}$'s are given by the following equations \cite{Garrard}.
\begin{equation}
\frac{\partial V_{0}^{T}}{\partial\pmb{\mathcal{X}}}\pmb{\mathcal{A}}\pmb{\mathcal{X}}- \frac{1}{4}\frac{\partial V_{0}^{T}}{\partial\pmb{\mathcal{X}}}\pmb{\mathcal{B}}\pmb{\widetilde{R}}^{-1}\pmb{\mathcal{B}}^{T}\frac{\partial V_{0}}{\partial \pmb{\mathcal{X}}} + \pmb{\mathcal{X}}^{T}\pmb{\widetilde{Q}}\pmb{\mathcal{X}} = 0,
\label{V0}
\end{equation}
and
\begin{equation}
\begin{split}
\frac{\partial V_{\mathfrak{n}}^{T}}{\partial\pmb{\mathcal{X}}}\pmb{\mathcal{A}}\pmb{\mathcal{X}} - \frac{1}{4}\frac{\partial V_{\mathfrak{n}}^{T}}{\partial\pmb{\mathcal{X}}}\pmb{\mathcal{B}}\pmb{\widetilde{R}}^{-1}\pmb{\mathcal{B}}^{T}\frac{\partial V_{0}}{\partial \pmb{\mathcal{X}}} \\ - \frac{1}{4}\frac{\partial V_{0}^{T}}{\partial\pmb{\mathcal{X}}}\pmb{\mathcal{B}}\pmb{\widetilde{R}}^{-1}\pmb{\mathcal{B}}^{T}\frac{\partial V_{\mathfrak{n}}}{\partial \pmb{\mathcal{X}}}  
+ \sum_{k=1}^{\mathfrak{n}-1}\frac{\partial V_{k}^{T}}{\partial \pmb{\mathcal{X}}} \pmb{f}_{\mathfrak{n}+1-k}\\ + \frac{\partial V_{0}^{T}}{\partial \pmb{\mathcal{X}}}\pmb{f}_{\mathfrak{n}+1}- \frac{1}{4}\sum_{k=1}^{\mathfrak{n}-1}\frac{\partial V_{k}^{T}}{\partial\pmb{\mathcal{X}}}\pmb{\mathcal{B}}\pmb{\widetilde{R}}^{-1}\pmb{\mathcal{B}}^{T}\frac{\partial V_{\mathfrak{n}-k}}{\partial \pmb{\mathcal{X}}} = 0.
\end{split}
\label{laborius}
\end{equation} 
The optimal control solution is then:
\begin{equation}
\pmb{\mathcal{U}} = -\frac{1}{2}\pmb{\widetilde{R}}^{-1}\pmb{\mathcal{B}}^{T}\sum_{\mathfrak{n}=0}^{\infty}\frac{\partial V_{\mathfrak{n}}}{\partial \pmb{\mathcal{X}}}
\label{optimal_control}
\end{equation}

It should be noted that $V_{0}$ is quadratic function in $\pmb{\mathcal{X}}$, $V_{1}$ is cubic in $\pmb{\mathcal{X}}$, and in general $V_{\mathfrak{n}}$ is of order $\mathfrak{n}+2$ in $\pmb{\mathcal{X}}$. Also, solution of $V_{\mathfrak{n}}$ leads to the $\mathfrak{n}+1$ order control term. Even though Eq. \ref{V0} is nonlinear, the solution is well known:
\begin{equation}
V_0 = \pmb{\mathcal{X}}^T \pmb{\widetilde{P}}\pmb{\mathcal{X}},
\end{equation}
where $\pmb{\widetilde{P}}$ is the solution of the algebraic Riccati equation
\begin{equation}
\widetilde{\pmb{P}}\pmb{\mathcal{A}}+\pmb{\mathcal{A}}^T \widetilde{\pmb{P}} - \widetilde{\pmb{P}} \pmb{\mathcal{B}}\widetilde{\pmb{R}}^{-1} \pmb{\mathcal{B}}^T \widetilde{\pmb{P}}=-\widetilde{\pmb{Q}}.
\label{Riccati}
\end{equation}
 Equation \ref{laborius} is linear, but in most cases cannot be solved analytically. Considering the augmented dynamical system with $n(P+1)$ states, the general procedure for obtaining approximate solutions is as follows \cite{Garrard}:

\begin{enumerate}
  \item Assume
  \begin{equation}
  \begin{split}
  V_{\mathfrak{n}} = \sum_{i_1=0}^{\mathfrak{n}+2} \cdots \sum_{i_{n(P+1)}=0}^{\mathfrak{n}+2} a^{\mathfrak{n}}_{i_1,\ldots,i_{n(P+1)}}x_{1}^{i_1}\cdots x_{n(P+1)}^{i_{n(p+1)}},\\
   {\rm s.t.} \ i_1+\cdots+i_{n(P+1)} \leq \mathfrak{n}+2;
  \end{split}
  \label{approxVpert}
  \end{equation}
  \item Calculate $\frac{\partial V_{\mathfrak{n}}}{\partial\pmb{\mathcal{X}}}$;
  \item Substitute $\frac{\partial V_{\mathfrak{n}}}{\partial\pmb{\mathcal{X}}}$ into Eq. \ref{laborius};
  \item Set the sum of coefficients of like terms equal to zero;
  \item Solve the resulting simultaneous linear algebraic equations for $a^{\mathfrak{n}}_{i_1,\ldots,i_{n(P+1)}}$.
\end{enumerate}

After $V_{\mathfrak{n}}$ is obtained, $\frac{\partial V_{\mathfrak{n}}}{\partial\pmb{\mathcal{X}}}$ can be calculated and substituted into Eq. \ref{optimal_control} to obtain $\pmb{\mathcal{U}}_{\mathfrak{n}+1}$. The convergence of this pertubational procedure and an estimate of the degradation of performance resulting from truncation is discussed in \cite{Garrard1969} and \cite{Lukes}.

\paragraph*{Remark 3.3.1} Note that the approximate solution of $V_{\mathfrak{n}}$ gives a suboptimal character to the control law in Eq. \ref{optimal_control}. Besides, the assumption in Eq. \ref{polynomialH} may not be the case for a general nonlinear dynamical system, and therefore the function $\pmb{\mathcal{H}}$ has to be approximated by a Taylor series expansion around the zero equilibrium point. Moreover, for practical reasons, the control law is Eq. \ref{optimal_control} has to be truncated.

\paragraph*{Remark 3.3.2} Optimality does not imply stability, unless the infinite horizon optimal control law $\pmb{\mathcal{U}}(\pmb{\mathcal{X}})$ is stabilizing, i.e., $\pmb{\mathcal{U}}(\pmb{0})=\pmb{0}$ and the zero solution of the closed-loop system in Eq. \ref{FinalPC2} is asymptotically stable in some neighborhood. Therefore, an assumption to apply the suboptimal control law derived in this section is that Eq. \ref{FinalPC2} is stabilizable \cite{Lukes,Markus}, that is, the Jacobian matrix
\begin{equation}
\pmb{\mathcal{J}}_{\pmb{0}} = \frac{\partial}{\partial \pmb{\mathcal{X}}} \left[ \pmb{\mathcal{A}}\pmb{\mathcal{X}}+ \pmb{\mathcal{H}}+\pmb{\mathcal{B}}\pmb{\mathcal{U}} \right]_{\pmb{\mathcal{X}}=\pmb{0}}
\end{equation}
is a stability matrix. That is the case if $\max [\rm{Re} (\lambda(\pmb{\mathcal{J}}_{\pmb{0}}))]<0$, where $\lambda(\pmb{\mathcal{J}}_{\pmb{0}})$ denotes the eigenvalues of the Jacobian matrix. Considering the control law in Eq. \ref{optimal_control}, the closed-loop dynamics is given by:
\begin{equation}
\pmb{\dot{\mathcal{X}}} = \pmb{\mathcal{A}}\pmb{\mathcal{X}}+\pmb{\mathcal{H}}-\frac{1}{2}\pmb{\mathcal{B}}\pmb{\widetilde{R}}^{-1}\pmb{\mathcal{B}}^{T}\sum_{\mathfrak{n}=0}^{\infty}\frac{\partial V_{\mathfrak{n}}}{\partial \pmb{\mathcal{X}}},
\label{closedXdot}
\end{equation}
and therefore the associated Jacobian matrix at the origin can be calculated by:
\begin{eqnarray}
\pmb{\mathcal{J}}_{\pmb{0}} &=& \frac{\partial}{\partial \pmb{\mathcal{X}}}\left[ \pmb{\mathcal{A}}\pmb{\mathcal{X}}+\pmb{\mathcal{H}}-\frac{1}{2}\pmb{\mathcal{B}}\pmb{\widetilde{R}}^{-1}\pmb{\mathcal{B}}^{T}\sum_{\mathfrak{n}=0}^{\infty}\frac{\partial V_{\mathfrak{n}}}{\partial \pmb{\mathcal{X}}} \right]_{\pmb{\mathcal{X}}=\pmb{0}}\\
&=&  \frac{\partial}{\partial \pmb{\mathcal{X}}}\left[ \pmb{\mathcal{A}}\pmb{\mathcal{X}}+\pmb{\mathcal{H}}-\frac{1}{2}\pmb{\mathcal{B}}\pmb{\widetilde{R}}^{-1}\pmb{\mathcal{B}}^{T}\frac{\partial V_{0}}{\partial \pmb{\mathcal{X}}} \right]_{\pmb{\mathcal{X}}=\pmb{0}},
\label{Jacobiancontrol0}
\end{eqnarray}
since $\frac{\partial V_{\mathfrak{n}}}{\partial \pmb{\mathcal{X}}}$ has a polynomial form in $\pmb{\mathcal{X}}$ of order $2$ or greater for $\mathfrak{n} \geq 1$. To determine under which conditions the Jacobian matrix in Eq. \ref{Jacobiancontrol0} is a stability matrix, consider the closed-loop system in Eq. \ref{closedXdot} with only linear control terms (i.e., the first term of the $V_{\mathfrak{n}}$ expansion only), and a Lyapunov function candidate of form:
\begin{equation}
\mathcal{V} = \pmb{\mathcal{X}}^T\widetilde{\pmb{P}}\pmb{\mathcal{X}}.
\label{LyapunovCan}
\end{equation}
Furthermore, define the following function:
\begin{equation}
L = \pmb{\mathcal{X}}^T \pmb{\widetilde{Q}} \pmb{\mathcal{X}}- \pmb{\mathcal{H}}^T \pmb{\widetilde{P}} \pmb{\mathcal{X}} - \pmb{\mathcal{X}}^T \pmb{\widetilde{P}} \pmb{\mathcal{H}}.
\label{Lpositivedefinite}
\end{equation}
 For positive definite function $L$ and positive definite matrix  $\widetilde{\pmb{R}}$, the derivative of the $\mathcal{V}$ function evaluated in the suboptimal trajectory is given by $\dot{\mathcal{V}}=-L-\pmb{\mathcal{U}}^T \pmb{\widetilde{R}}\pmb{\mathcal{U}}$ and it is negative definite. Then $\mathcal{V}$ is a Lyapunov function and the controlled system is locally asymptotically stable. Hence the zero equilibrium point in the closed loop dynamical system with only linear control terms is asymptotically stable if the function in Eq. \ref{Lpositivedefinite} is positive definite. Consequently, under such condition, the Jacobian matrix in Eq. \ref{Jacobiancontrol0} must be a stability matrix.  Note that the Jacobian matrix at the origin for the closed-loop system with linear and nonlinear control are the same. Therefore, the nonlinear control in Eq. \ref{optimal_control} is stabilizing and the closed-loop trajectories initiated within a neighborhood of the origin asymptotically converge to the equilibrium point. Note that this is a local result. Global stability characteristics should be studied case by case via Lyapunov's direct method.

\subsection{Implementation}
\label{Implementation}
In Sections \ref{Optimal Control Design} \ref{Nonlinear Suboptimal Control}, a nonlinear optimal control strategy was presented. The method was derived for the augmented dynamical system for the gPC mode strengths, that is, the control laws map $\pmb{\mathcal{X}}$ to $\pmb{\mathcal{U}}$. In an actual application, however, the mode strengths are not physical quantities, thus they cannot be measured. Therefore, the gPC coefficients have to be estimated in real time, as well as the uncertain parameters, in order to allow the computation of the optimal control $\pmb{\mathcal{U}}$ and the subsequent determination of the actual control input $\pmb{u}(t,\pmb{\xi})$ for the dynamical system, using Eq. \ref{gPCu}. From Section \ref{Polynomial Chaos}, we know that the gPC mode strengths of the decomposition of $\pmb{x}(t,\pmb{\xi}) \in \mathbb{R}^n$ are given by Eq. \ref{PCEgcoeff}, which, for a truncated decomposition, can be written in a more convenient way:
\begin{equation}
x_{i,k} =E[x_i(t,\pmb{\xi})]~~~~~~~~~~\mbox{for }~i=1,\ldots,n ~~~\mbox{and }~k=0,
\label{xcov22}
\end{equation}
and
\begin{equation}
x_{i,k} = \frac{\mbox{cov}[x_i(t,\pmb{\xi}),\Psi_k (\pmb{\xi})]}{V[\Psi_k (\pmb{\xi})]}~~~~~\mbox{for }~i=1,\ldots,n ~\mbox{and }~k=1,\ldots,P.
\label{xcov}
\end{equation}
Note that, in Eq. \ref{xcov}, the variance $V[\Psi_k (\pmb{\xi})]$ can be computed off-line since the polynomial basis is known for a given set of uncertain parameters. Therefore, the determination of the gPC mode strengths of the actual state vector boils down to the computation of the covariance between the states and the polynomial basis at a certain time instant. The estimation of the set of uncertain parameters $\pmb{\xi} \in \mathbb{R}^d$ can be performed through different on-line parameter estimation algorithms, such as the recursive least square method or the maximum likelihood technique. \cite{Ljung} present several methods for recursive parameter estimation.  We denote the estimated parameters by $\hat{\pmb{\xi}}$. Each parameter in the vector $\pmb{\xi}$ is assumed to be time-invariant, that is, each overcome of the random variables are constant over time. Therefore, the parameter estimation algorithm has to run only until all parameters have been estimated, when the gain of the algorithm has ultimately decreased to zero \cite{Goodwin}. The convergence of the parameter estimation, however, is dependent on some properties of the input signals in the dynamical system. As discussed by \cite{Narendra}, this signal property is referred to as persistent excitation (PE) and it is pivotal for on-line parameter estimation. Note that this usually implies that the control input cannot be monotone in order to excite the plant, ergo antagonistic toward the regulation objective. As a consequence, there is a trade-off between the estimation and the regulation performances.

\begin{figure*}[h!]
\centering
\scalebox{0.8}{\begin{tikzpicture}
\tikzstyle{myedgestyle} = [-triangle 60]
\draw [dashed, fill=gray!20] (-5,3.5) rectangle (4.1,-2);
\draw [dashed, fill=gray!50]  (4.75,1) rectangle (8.65,-4.5);

\node at (1.1,0.75){$\Psi_{k}(\pmb{\xi})$};
\draw[-triangle 60](5,0) rectangle(8,-1.5);
\draw[-triangle 60](5,-2.5) rectangle(8,-4);
\draw[-triangle 60](10.5,0) rectangle(13.5,-1.5);
\draw[-triangle 60](7,3) rectangle(10,1.5);

\draw[-triangle 60](3.5,2.3) --(7,2.3);
\draw[-triangle 60](10,2.3) --(13.5,2.3);
\draw[-triangle 60](12,2.3) --(12,0);
\draw[-triangle 60](10.5,-0.75) --(8,-0.75);
\draw[-triangle 60](5,-0.75) --(3.5,-0.75);
\draw[-triangle 60](9.5,-0.75) --(9.5,-3.25)--(8,-3.25);
\draw[-triangle 60](5,-3.25) --(-4.55,-3.25)--(-4.55,2.3)--(-4.15,2.3);
\node at (-2.5,-0.8){Controller};
\node at (5,2.6){$u_{i}(t,\pmb{\xi})$};
\node at (11,2.6){$x_{i}(t,\pmb{\xi})$};
\node at (8.5,2.5){Dynamical};
\node at (8.5,2.0){system};
\node at (12,-0.75){$\pmb{c}(\cdot)$};
\node at (6.5,-0.75){Estimation of $\pmb{\xi}$};
\node at (6.5,-3.25){Estimation of $\pmb{\mathcal{X}}$};
\node at (4.45,-0.5) {$\hat{\pmb{\xi}}$};
\node at (4.45,-3) {$\hat{\pmb{\mathcal{X}}}$};
\node at (9.85,-0.5){$y$};
\draw  (-4.15,3) rectangle (-1,1.5);
\draw  (3.5,3) rectangle (0.5,1.5);
\draw  (3.5,0) rectangle (0.5,-1.5);
\draw [-triangle 60] (-1,2.3)-- (0.5,2.3);
\draw [-triangle 60] (2,0) -- (2,1.5);
\node at (-2.55,2.7) {Linear or};
\node at (-2.55,2.3) {nonlinear control};
\node at (-2.55,1.9) {law};
\node at (2,2.3) {$\sum_{k=0}^{P} u_{i,k}\Psi_{k}(\pmb{\xi})$};
\node at (-0.25,2.6) {$\pmb{\mathcal{U}}$};
\node at (2,-0.75) {$\Psi_{k}(\cdot)$};
\node at (6.7,0.5) {Estimation algorithm};

\end{tikzpicture}}
\caption{Block Diagram.}
\label{fig7} 
\end{figure*}
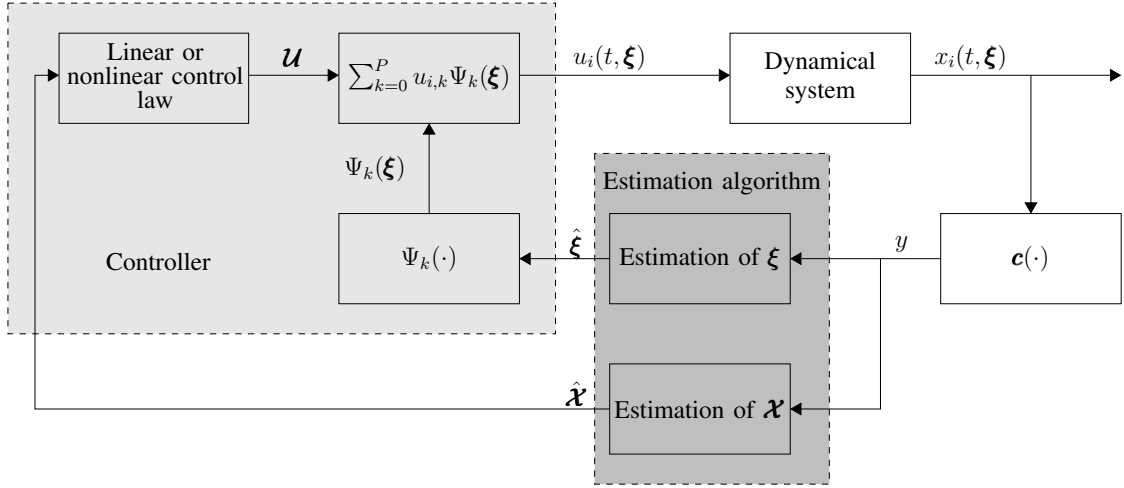

Once the polynomial bases are totally determined by the parameter estimation, state estimation will be required to compute the covariance in Eq. \ref{xcov}. The state estimation algorithm, which is required for full state feedback control of stochastic systems, can provide estimates of the covariance matrix in real time, inasmuch as the algorithm keeps track of the probability density function of the state. In this work, we consider on-line estimation algorithms for discrete measurements of continuous dynamics, that is, the algorithm deals with sequential data, which requires that the states and parameter estimates be recursively updated within the the sampling period time \cite{Goodwin}. We denote the states estimates by $\hat{\pmb{x}}$. Figure \ref{fig7} shows a scheme of the implementation of the polynomial-chaos based control strategy, where the measurements $\pmb{y} \in \mathbb{R}^s$ are given by: 
\begin{equation}
\pmb{y} = \pmb{c}(\pmb{x}).
\end{equation}
Note that the proposed approach bears a resemblance to traditional adaptive control strategies, since the model is not completely specified and we combine on-line parameter estimation with on-line control, providing a self-learning nature to the controller \cite{Goodwin}. The remarkable difference between the traditional adaptive control and the polynomial chaos-based control is that the latter takes the uncertain parameters as random variables and incorporates the probability density function of the state in the feedback law.

In this work we suggest Kalman filtering for adaptive state estimation, that is, the states and parameters are estimated simultaneously. To implement the adaptive state estimation of Eq \ref{StochSys}, we define an augmented state of the form (not to be confused with the augmented system for the gPC coefficients):
\begin{equation}
\pmb{\underaccent{\bar}{x}}(t) \triangleq \left[ \begin{array}{c} \pmb{x}(t,\pmb{\xi}) \\ \pmb{\xi} \end{array} \right] \in \mathbb{R}^{n+d},
\end{equation}
such that
\begin{equation}
\dot{\pmb{\underaccent{\bar}{x}}}(t) = \pmb{\underaccent{\bar}{f}}(\pmb{\underaccent{\bar}{x}}(t),\pmb{u}(t)) + \pmb{\underaccent{\bar}{G}}\pmb{\omega}(t)
\label{xundbar}
\end{equation}
with measurements at the discrete time $k+1$
\begin{equation}
\pmb{\underaccent{\bar}{y}}_{k+1} = \pmb{c}_{k+1}(\pmb{\underaccent{\bar}{x}}_{k+1})+\pmb{\nu}_{k+1},
\end{equation} 
where
\begin{equation}
\pmb{\underaccent{\bar}{f}}(\pmb{\underaccent{\bar}{x}}(t),\pmb{u}(t)) \triangleq \left[ \begin{array}{c} \pmb{A}(t,\pmb{\xi})\pmb{x}(t,\pmb{\xi})+\pmb{h}(t,\pmb{x},\pmb{\xi})+\pmb{B}(\pmb{\xi})\pmb{u}(t,\pmb{\xi}) \\ \pmb{0}_{d\times 1} \end{array} \right]  \in \mathbb{R}^{n+d},
\end{equation}
$\pmb{\underaccent{\bar}{G}} \in \mathbb{R}^{n\times n_w}$, and the signal $\{\pmb{\omega} (t)\} \in \mathbb{R}^{n_w}$ and the sequence $\{ \pmb{\nu}_{k} \} \in \mathbb{R}^s$ are noise terms, which are assumed to be realizations of zero mean white Gaussian processes $\{ \pmb{W}(t) \}$ and $\{ \pmb{N}_{k} \}$, respectively, such that $\pmb{W}(t) \sim (\pmb{0}, \pmb{Q_{\omega}}(t))$ and $ \pmb{N}_{k} \sim (\pmb{0},\pmb{R}_{\pmb{\nu}k})$, where $\pmb{Q_{\omega}}(t) \in \mathbb{R}^{n_w \times n_w}$ and $\pmb{R}_{\pmb{\nu}k} \in \mathbb{R}^{s\times s}$ are known covariance matrices. The initial condition of \ref{xundbar}, $\pmb{\underaccent{\bar}{x}}_0$, is an outcome of the random variable $\pmb{\underaccent{\bar}{X}}_0 \sim (\bar{\pmb{\underaccent{\bar}{x}}},\bar{\pmb{\underaccent{\bar}{P}}})$, where $\bar{\pmb{\underaccent{\bar}{x}}} \in \mathbb{R}^n$ and $\bar{\pmb{\underaccent{\bar}{P}}} \in \mathbb{R}^{n\times n}$ are known. By assumption, $\pmb{\underaccent{\bar}{X}}_0$, $\{ \pmb{W}(t) \}$ and $\{ \pmb{N}_k \}$ are mutually uncorrelated. The addition of the noise signals accounts for the unmodeled disturbances and uncertainties in the environment and model in a real application. Note that we consider that the uncertain parameters $\pmb{\xi}$ are constant over time, that is, we estimate an output of the random variables. From a theoretical standpoint, there are no restrictions on the number of unknown parameters in $\pmb{\xi}$. In practice, however, the adaptive filter might encounter observability issues for a large $\pmb{\xi}$, and the user may want to explore other estimation strategies.    

Note that, in Eq. \ref{xcov}, the covariance between the state $x_i(t,\pmb{\xi})$ and the polynomial basis $\Psi_k (\pmb{\xi})$ can be rewritten as a higher order moment between the state and the vector of parameters $\pmb{\xi}$. In general, to compute the gPC mode strength $x_{i,k}$, it is necessary moments of order $k+1$ between the state and $\pmb{\xi}$. The adaptive state estimation using the extended Kalman filter (EKF) only provides the first and second moments. Therefore, for a polynomial chaos expansion with $P>1$, numerical approximations of high-order moments are required. This could be a drawback of the implementation of the polynomial chaos-based control using adaptive estimation with EKF, since the computational cost to estimate such moments might be significant. There are several methods to estimate high-order moments in the literature \cite{Dutta,Liu,Ponomareva,Grothe}. For instance, \cite{Majji} present the so called Jth Moment Extended Kalman Filter, a estimation framework for high order moments of nonlinear dynamical systems using the state transition tensor approach. The implementation of this method for vector models is usually quite laborious due to the algebraic transformations involved resulting in a high computation expenditure.


Next we present a summary for the control design and implementation of the method proposed in this work:
\begin{itemize}
\item
Step 1: Given a dynamical system of form as in Eq. \ref{StochSys}, with $r$ uncertain parameters with known distribution, determine the expansion basis according to Wiener-Askey scheme in Tab. \ref{tab:1}.

\item
Step 2: Choose the polynomial order $d$ so that the expansion error is small. The gPC will then have $P=\frac{(r+d)!}{r!d!}-1$ terms.

\item
Step 3: Construct the augmented system as shown in Section \ref{sec:1.2}. One should obtain a deterministic system as in Eq. \ref{FinalPC}. Once the augmented system is determined, one can analyze the first two statistical moments of the model by using Eq. \ref{PCE}. Such results can be compared with Monte Carlo (MC) simulations of the original stochastic system. The error between the results obtained through gPC and the results from MC should be small. Otherwise, one should return to Step 2 and increase $d$.   

\item
Step 4: Choose a linear or nonlinear control strategy as presented in Section \ref{Optimal Control Design}. As for the nonlinear control, one has to solve the HJB equation by using the perturbational procedure with expansion in Eq. \ref{PowerSeries} truncated at order $\mathfrak{n}$. Obtain the nonlinear control of order $\mathfrak{n}+1$ using Eq. \ref{optimal_control}. To do so, solve Eq. \ref{V0} and \ref{laborius} using the approximation in Eq. \ref{approxVpert}. For any control law designed, the control weights $\pmb{Q}$ and $\pmb{R}$ have to be chosen so that stability criterion is met.  One can verify such condition by simulating the closed loop augmented system and checking if the function in Eq. \ref{Lpositivedefinite} is always positive definite.

\item
Step 5: Implement the closed loop system as shown in Fig. \ref{fig7}. An estimation algorithm should be chosen to provide online estimates of the states and the uncertain parameters. The gPC mode strengths can be computed using Eq. \ref{xcov22} and \ref{xcov}. For $P=1$, an Extended Kalman Filter or any other formulation of the Kalman Filter for nonlinear estimation should suffice. For $P>1$, one should implement an estimation algorithm that provides information about higher order moments.

\end{itemize}

\section{Aircraft Model}
\label{Aircraft Model}
\begin{figure}[h!]
\centering
\includegraphics[width=0.95\linewidth]{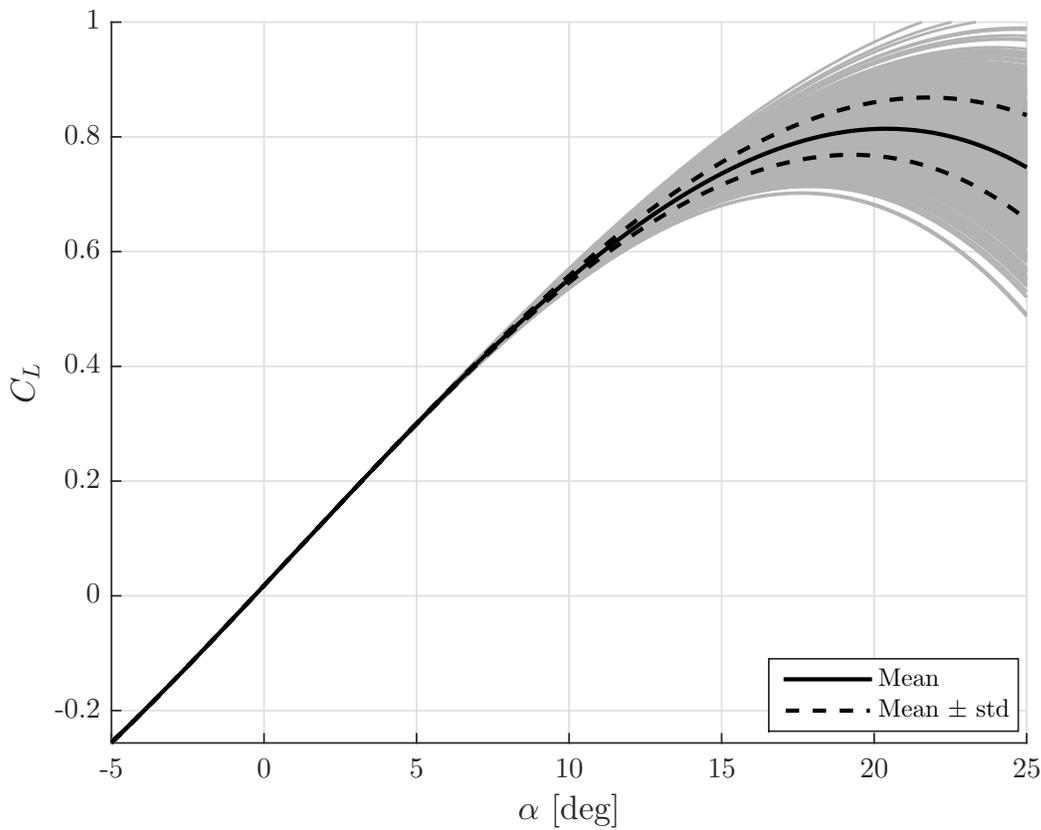}
\caption{Lift coefficient with uncertainty.}
\label{fig6} 
\end{figure}

 This work is concerned with the longitudinal dynamics of  fighter aircraft,  focusing on control design for high-angle-of-attack flights. Such flight regime is critical for high performance combat aircraft in rapid maneuvers such as evasion, pursuit, and nose pointing to obtain the first opportunity of firing the weapons \cite{Atesoglu}. Examples are the Cobra and Herbst maneuvers, in which the aircraft for short periods has to attain high angular velocities at extreme angles of attack. If we consider level flight with zero sideslip angle and no wind disturbances, and assuming that the aircraft flies at constant velocity, the nonlinear model for the longitudinal motion is given by the following set of equations:
\begin{equation}
\begin{split}
\dot{\alpha} = -\frac{1}{{\rm m u}}\left[q_{\infty} S \cos ^3(\alpha ) C_L -{\rm m} g \cos ^2(\alpha ) \cos (\theta ) \right. \\ \left.-{\rm m u} \dot{\theta} \cos ^2(\alpha )-2 q_{\infty} S \sin (\alpha ) \cos ^2(\alpha )
   C_D \right. \\ \left. -{\rm m}
   {\rm u}\dot{\theta} \sin ^2(\alpha ) -{\rm m} g \sin (\alpha ) \cos (\alpha ) \sin (\theta ) \right. \\ \left.    -q_{\infty} S \sin ^2(\alpha ) \cos (\alpha ) C_L\right],
   \end{split}
   \label{eqAlpha}
\end{equation}
\begin{equation}
\dot{\theta} = q,
\label{eqTheta}
\end{equation}
\begin{equation}
\dot{q} = \frac{1}{I_{yy}}c q_{\infty} S C_M.
   \label{eqQ}
\end{equation}
where $\alpha$ is the angle of attack, $\theta$ is the pitch angle, $q$ is the pitch rate, $q_{\infty}$ is the dynamic pressure, $g$ is the acceleration of gravity, ${\rm u}$ is the aircraft speed, $S$ is the wing area, $c$ is the mean aerodynamic chord, and ${\rm m}$ and $I_{yy}$ are the mass and moment of inertia of the vehicle, respectively. The terms $C_L$, $C_D$ and $C_M$ are aerodynamic coeffcients for lift, drag and moment, respectively. We consider a blended-wing-body fighter aircraft whose parameters are shown in Tab. \ref{tab4}.
\begin{table}[h!]
\caption{Aircraft parameters.}
\label{tab4}

\center
\begin{tabular}{cccc}
  \hline
	Parameter & Description & Value & Unit  \\
	\hline
  ${\rm m}$ & 	Mass &  8,780  & kg\\
  $I_{yy}$ & Moment of inertia     &  13,418 & kg$\cdot$m$^2$\\
	$S$ & Wetted area  &  50.2 & m$^2$\\
	$c$ & Mean aerodynamic chord & 4.12 & m\\
	$b$ & Wing span & 15.84 & m \\
	$V_c$ & Cruise speed & 171.3 & m/s \\
	$h_c$ & Cruise altitude & 30,000& m \\
	\hline
\end{tabular}
\end{table}
The lift coefficient typically varies linearly with the angle of attack at low to moderate $\alpha$. In this region the airflow moves smoothly over the aircraft surface and is attached over most part of it. At high angles of attack, however, the flow tends to separate from the top surface of the airfoil due to viscous effects creating a large wake behind the airfoil and a condition of reversed flow. As a consequence, the lift force is decreased while the drag grows appreciably \cite{Anderson}. Under such conditions, the aircraft is said to be stalled, and the lift coefficient is no longer a linear function of $\alpha$. The drag and moment coefficient are also nonlinear functions of $\alpha$. To fully characterize the behavior of the aerodynamic coefficients as a nonlinear function of the angle of attack, data are usually obtained experimentally in wind-tunnel tests. For instance, the solid black line in Fig. \ref{fig6} shows the lift coefficient obtained experimentally for the combat aircraft considered.

To incorporate these coefficients to the mathematical model in Eq \ref{eqAlpha} - \ref{eqQ}, a third order polynomial is fitted to the experimental data. This approximation is usual in the literature of dynamics and control of high-angle-of-attack flights, such as in the works of \cite{Garrard,Liaw,PereiraBaltha}. Such polynomials have the following standard form:
\begin{equation}
C_{(~)} = C_{(~)0}+C_{(~)1}\alpha +C_{(~)2}\alpha^2 +C_{(~)3}\alpha^3 + \Delta C_{(~)},
\label{eqcoeff} 
\end{equation}
where $C_{(~)0}, C_{(~)1}, C_{(~)2}$ and $C_{(~)3}$ are the polynomial coefficients, and $\Delta C_{(~)}$ represent terms not directly related to $\alpha$. Table \ref{tab2} shows the coefficients that approximate the aerodynamic data obtained in wind tunnel tests for the fighter aircraft.
\begin{table}[h!]
\caption{Third order polynomial coefficients for aerodynamic curves}
\label{tab2}

\center
\begin{tabular}{ccccc}
  \hline
	Coefficient & $C_{(~)0}$ & $C_{(~)1}$& $C_{(~)2}$&$C_{(~)3}$  \\
	\hline
  $C_L$ & $0.0179$ &  $3.2569$  & $0.5450$ &$-9.6098$\\
  $C_D$ & $0.0355$     &  $- 0.1171$ & $1.6552$&$0.8908$\\
	$C_M$ & $- 0.0332$  &  $- 2.8543$ & $0.8669$&$2.3927$\\
	\hline
\end{tabular}
\end{table}

Even though the third order polynomials provide a reasonable approximation for the aerodynamic coefficients at low to high angles of attack, there may still be nonlinear phenomena associated with the stall condition that were not fully captured in the wind-tunnel tests. As described by Rom \cite{Rom}, the stall evolution is a complicated process due to vortex breakdown, which leads to the flow braking up into a nonsteady turbulent wake. Besides the complicated stall process that might not be fully characterized by the function approximation, the data obtained in the wind-tunnel tests may not be completely accurate since there are similarity parameters not accounted for \cite{Anderson}. Moreover, the final validation of the mathematical model is performed in flight tests, which are usually not completed before the first control design routine. For these reasons, there may be a mismatch between the real aerodynamic coefficient curves and the mathematical model, due to unmodeled phenomena, especially in high angles of attack, where the nonlinearites are more significant. To account for that, we add a parameter of uncertainty in the lift coefficient, such that Eq. \ref{eqcoeff} for this coefficient is now written as:
\begin{equation}
C_{L} = C_{L0}+C_{L1}\alpha +C_{L2}\alpha^2 +(C_{L3}+\xi)\alpha^3 + \Delta C_{L},
\label{CLrnd}
\end{equation}
where $\xi \sim \mathcal{N}(0,\sigma_{C_L}^{2})$ is a normal random variable. Figure \ref{fig6} shows in grey a Monte Carlo simulation of Eq. \ref{CLrnd} with 1000 outcomes, as well as the standard deviation around the nominal lift coefficient obtained experimentally. We choose to add only one parameter of uncertainty in order to keep this example simple for the reader's benefit, and to avoid the high computational cost and observability issues that multiple unknown parameters can have.

To complete the longitudinal aerodynamic model, the increments $\Delta C_{(~)}$ in Eq. \ref{eqcoeff} are modeled as linear functions of the control surface deflections and the linear and angular velocities and accelerations \cite{Lewis}. For lift and moment increments we have:
\begin{equation}
\Delta C_{L} = C_{L_{q}}\frac{c}{2V_c}\dot{\theta} + C_{L_{u}}\frac{V_t}{V_c}+C_{L_{\delta_e}}\delta_e,
\label{deltacl}
\end{equation}
\begin{equation}
\Delta C_{M} = C_{M_{q}}\frac{c}{2V_c}\dot{\theta} + C_{M_{u}}\frac{V_t}{V_c}+C_{M_{\delta_e}}\delta_e.
\label{deltacm}
\end{equation}
where $V_t = \sqrt{\rm{u}^2 + \rm{v}^2 + \rm{w}^2}$ is the total velocity and $V_c$ is the cruise speed. The elevon deflection is represented by $\delta_e$. The coefficients in Eq. \ref{deltacl} and \ref{deltacm} are shown in Tab.\ref{tab3}
\begin{table}[h!]
\caption{Aerodynamic coefficients}
\label{tab3}
\center
\begin{tabular}{cccc}
  \hline
	Coefficient & $C_{(~)_q}$ & $C_{(~)_u}$& $C_{(~)_{\delta_e}}$   \\
	\hline
  $C_L$ & $4.649$ &  $0.0919$  & $-0.2532$ \\
  $C_M$ & $-0.9064$     &  $0$ & $-4.599$\\
	\hline
\end{tabular}
\end{table}
Increments in the drag coefficient are not significant and therefore are not considered in this work. Note that the elevon deflection $\delta_e$ is related to pitch control and thus is the input commands. For control design purposes, following the derivations presented in Section \ref{Control Design}, these equation are conveniently rewritten in the state space form as follows:
\begin{equation}
\dot{\pmb{x}}(t,\xi) = \pmb{A}\pmb{x}(t,\xi)+\pmb{H}(\pmb{x},\xi)+\pmb{B}u(t)
\label{Dinamica}
\end{equation} 
While Eq. \ref{Dinamica} is used for control design, we consider a more general model for simulation, which includes state and sensor noise.
Therefore, for an outcome of random variable $\xi$ in Eq. \ref{Dinamica}, we have:
\begin{equation}
\dot{\pmb{x}}(t) = \pmb{A}\pmb{x}(t)+\pmb{H}(\pmb{x},t)+\pmb{B}u(t) + \pmb{G}\omega(t),
\label{dotXstoch}
\end{equation}
with measurements at the discrete time $k+1$
\begin{equation}
\pmb{y}_{k+1} = \pmb{C}\pmb{x}_{k+1}+\pmb{\nu}_{k+1},
\label{ModelMeasure}
\end{equation}
where 
\begin{equation}
\pmb{G} \triangleq  \left[ \begin{array}{c} 0\\0\\1 \end{array} \right] \in \mathbb{R}^3,
\end{equation}
\begin{equation}
\pmb{C} \triangleq  \pmb{I}_3 \in \mathbb{R}^{3\times 3},
\end{equation}
and the signal $\{\omega (t)\} \in \mathbb{R}$ and the sequence $\{ \pmb{\nu}_{k} \} \in \mathbb{R}^3$ are noise terms, which are assumed to be realizations of zero mean white Gaussian processes $\{ W(t) \}$ and $\{ \pmb{N}_{k} \}$, respectively, such that $W(t) \sim (0, Q_{\omega})$ and $ \pmb{N}_{k} \sim (\pmb{0},\pmb{R_{\nu}})$, where $Q_{\omega} \in \mathbb{R}$ and $\pmb{R_{\nu}}\in \mathbb{R}^{3\times 3}$ are known covariance matrices. The initial condition of \ref{dotXstoch}, $\pmb{x}_0$, is an outcome of the random variable $\pmb{X}_0 \sim (\bar{\pmb{x}},\bar{\pmb{P}})$, where $\bar{\pmb{x}} \in \mathbb{R}^3$ and $\bar{\pmb{P}} \in \mathbb{R}^{3\times 3}$ are known. By assumption, $\pmb{X}_0$, $\{ W(t) \}$ and $\{ \pmb{N}_k \}$ are mutually uncorrelated.

\section{Simulations}
\label{Simulations}
In the numerical simulations performed, we considered the longitudinal model presented in Section \ref{Aircraft Model}, with parameters given in Tab. \ref{tab4}, \ref{tab2} and \ref{tab3}. Here, it is important to calculate the trim condition, that is, the angle of attack and the elevon deflection during cruise flight. To calculate such condition, we find the equilibrium point of Eq. \ref{eqAlpha}-\ref{eqQ}, recalling that, for a symmetric airplane with leveled wings and zero sideslip angle, the following relation holds: $\theta_T = \alpha_T +\gamma$, where subscript $T$ denotes trim and $\gamma$ is the climb angle. For $\gamma = 0$ we have
\begin{eqnarray}
\alpha_T &=& 2.47 \mbox{ deg},\\
\delta_{eT} &=& -1.92 \mbox{ deg}.
\end{eqnarray}
The gains used in the control derivations, as shown in Sections \ref{Optimal Control Design} and \ref{Nonlinear Suboptimal Control}, were selected as:
\begin{eqnarray}
\pmb{Q} &=& \left[ \begin{array}{ccc} 1&0&0\\0&1&0\\0&0&1 \end{array}\right],\\
R &=& 1.
\end{eqnarray}
Linear, second, third and fifth order controllers were computed. Note that the gains were kept the same for all control designs in order to make a fair comparison among the performance of the controllers. In this work, due to the computational cost, we will only consider the implementation of control laws in an adaptive structure, as shown in Fig. \ref{fig7}, with gPC expansions up to $P=1$. Therefore, the adaptive state estimation with the EKF, or any other formulation of the nonlinear Kalman filter, can be used to obtain on-line estimates of the gPC mode strengths. In view of the highly nonlinear dynamical model for the aircraft considered in Section \ref{Aircraft Model}, we suggest the Unscented Kalman filter (UKF) in lieu of EKF.
Considering the full simulation model in Eq. \ref{dotXstoch} and \ref{ModelMeasure}, the following parameters were used in all simulations:
\begin{eqnarray}
Q_{\omega} &=& 0.1745, \\
\pmb{R_{\nu}} &=& \left[ \begin{array}{ccc} 0.81&0&0\\0&0.81&0\\0&0&0.81 \end{array}\right]\times 10^{-4}, \\
\bar{\pmb{P}} &=& \left[ \begin{array}{ccc} 3&0&0\\0&3&0\\0&0&3 \end{array}\right],\\
\bar{\pmb{x}} &=& \left[ \begin{array}{c} \alpha_0 -0.0087 \mbox{ rad}\\0 \mbox{ rad}\\0.0261 \mbox{ rad/s} \end{array}\right]. \\
\end{eqnarray}
As for the augmented system for the adaptive Kalman Filter algorithm presented in Section \ref{Implementation}, the following parameters were used:
\begin{eqnarray}
\bar{\pmb{\underaccent{\bar}{P}}} &=& \left[ \begin{array}{cc} \bar{\pmb{P}}&\pmb{0}_{3 \times 1}\\ \pmb{0}_{1 \times 3}&30 \end{array}\right],\\
\bar{\pmb{\underaccent{\bar}{x}}} &=& \left[ \begin{array}{c} \bar{\pmb{x}} \\ -9.6098 \end{array}\right]. 
\end{eqnarray}
where $\alpha_0$ denotes the initial condition for the angle of attack. Two initial conditions were simulated for each control law: $\pmb{x} = [\begin{array}{ccc}25^{o}&0&0 \end{array}]^T$ and $\pmb{x} = [\begin{array}{ccc}30^{o}&0&0 \end{array}]^T$. All initial conditions correspond to angles of attack within the stall region, i.e., greater than 20 degrees. 

\begin{figure}[h!]
\centering
\includegraphics[width=1.1\linewidth]{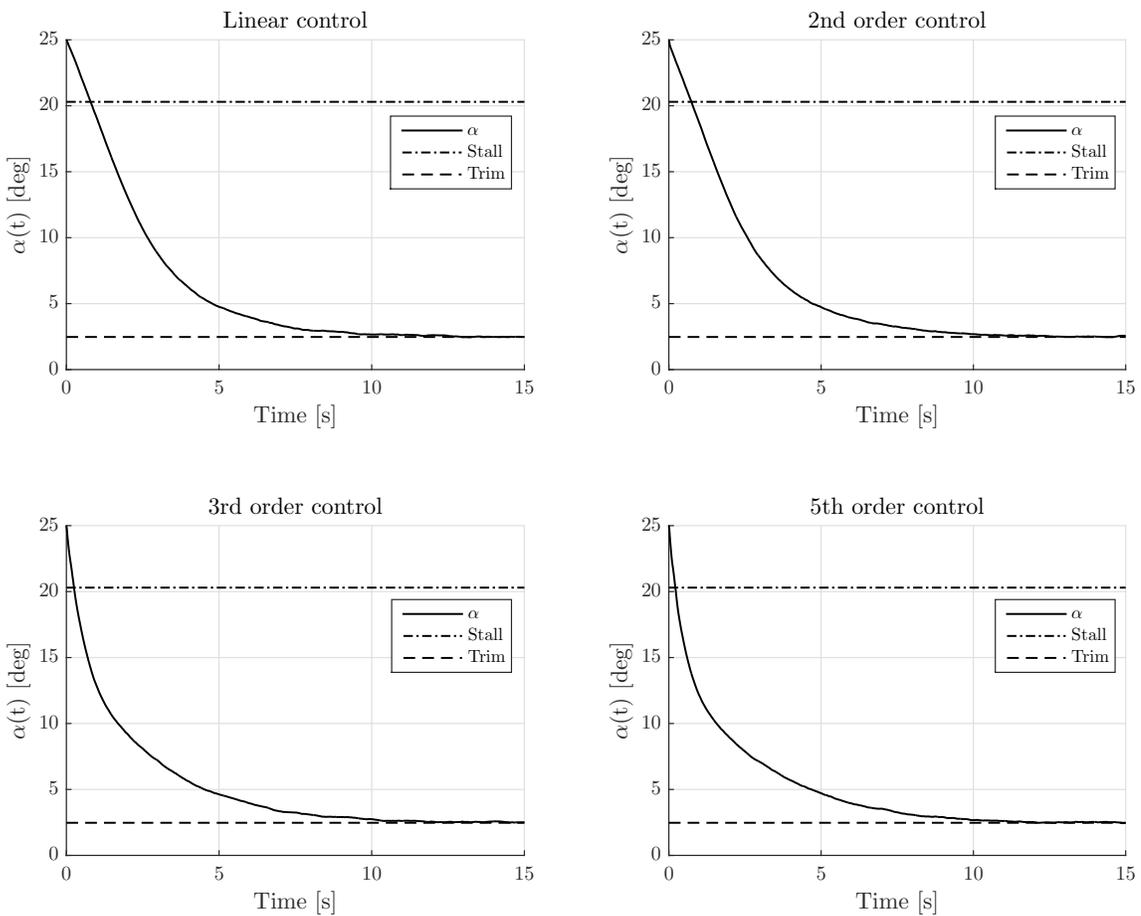}
\caption{Angle of attack of the aircraft with different control strategies. Initial condition $\alpha = 25$ deg.}
\label{fig12} 
\end{figure}

\begin{figure}[h!]
\centering
\includegraphics[width=1.1\linewidth]{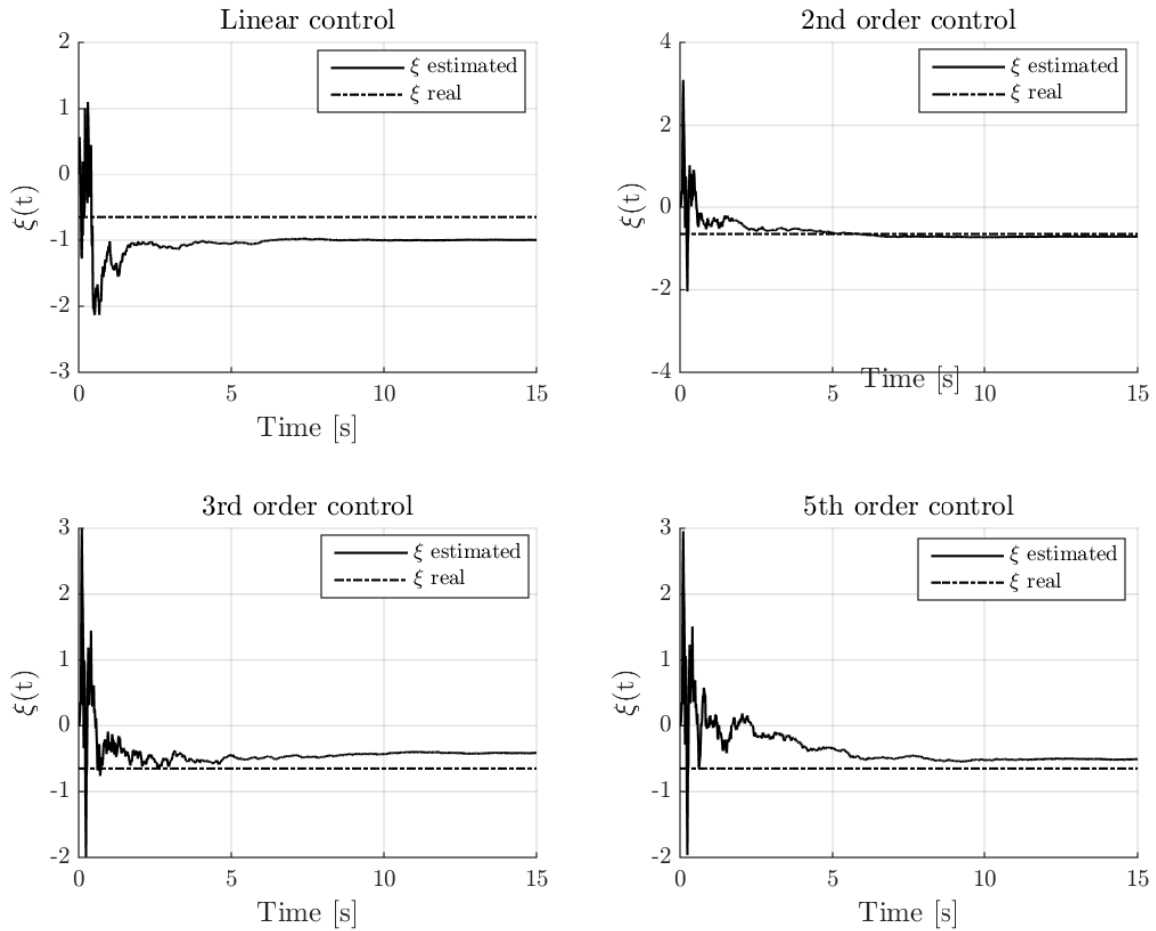}
\caption{Parameter estimation for the aircraft with different control strategies. Initial condition $\alpha = 25$ deg}
\label{fig16} 
\end{figure}

\begin{figure}[h!]
\centering
\includegraphics[width=1.1\linewidth]{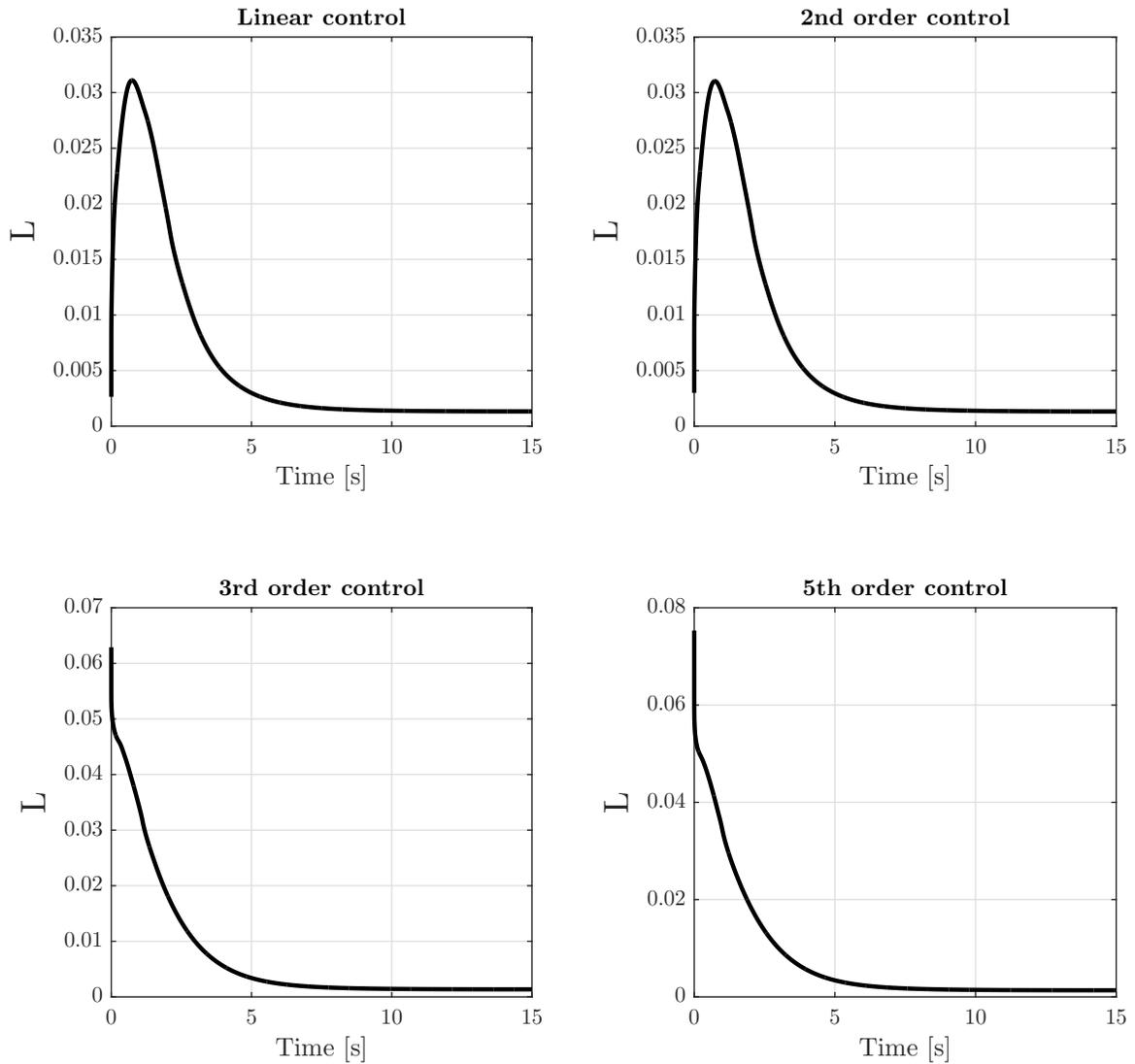}
\caption{Stability criterion verification for the aircraft with different control strategies. Initial condition $\alpha = 25$ deg}
\label{fig17} 
\end{figure}

In the first simulation, we consider $\alpha_0 = 25$ deg, which corresponds to an angle of attack within the stall region but still with little loss of lift. Figure \ref{fig12} shows the angle of attack response for each controller. In every simulation, the aircraft was successfully recovered from stall and reached the trim condition in approximately 10s. The 3rd and 5th order controllers provided a faster response in bringing the vehicle away from the stall condition. 
Figure \ref{fig16} shows the parameter estimation $\xi$ provided by the adaptive UKF, which is crucial for computing the actual control signal. We see that the estimation did not converge to the real value, but nevertheless it provided a reasonable approximation. The parameter estimation for the system with the 2nd order control had the better performance, whereas the estimation for the system with linear control had the worst performance.  Such issue can be explained in two ways: either the control input was not sufficiently PE or the parameter had minor influence in the dynamics. In the first case, as discussed in Section \ref{Implementation}, the convergence of parameter estimation is strongly dependent on whether the control signal is PE. In the second case, since the uncertainty is more intense at high angles of attack, as shown in Fig. \ref{fig6} an initial condition at $\alpha_0 = 25$ deg may not have been high enough to provide sufficient information for estimating $\xi$ accurately. 
Lastly, we check the the stability criterion for the closed-loop system with each controller. As discussed in Section \ref{Nonlinear Suboptimal Control}, the origin of the system is asymptotically stable in probability if the function $L$, in Eq. \ref{Lpositivedefinite} is positive definite. Figure \ref{fig17} shows the function for each control strategy. In all cases, $L$ is positive during all time instants and therefore we infer the stability in probability of the equilibrium point in the neighborhood that contains the initial condition. One should note that we verify the stability for the closed-loop stochastic system by analyzing a deterministic function for the gPC mode strengths with low computational cost when compared to traditional methods based on sampling.

\begin{figure}[h!]
\centering
\includegraphics[width=1.1\linewidth]{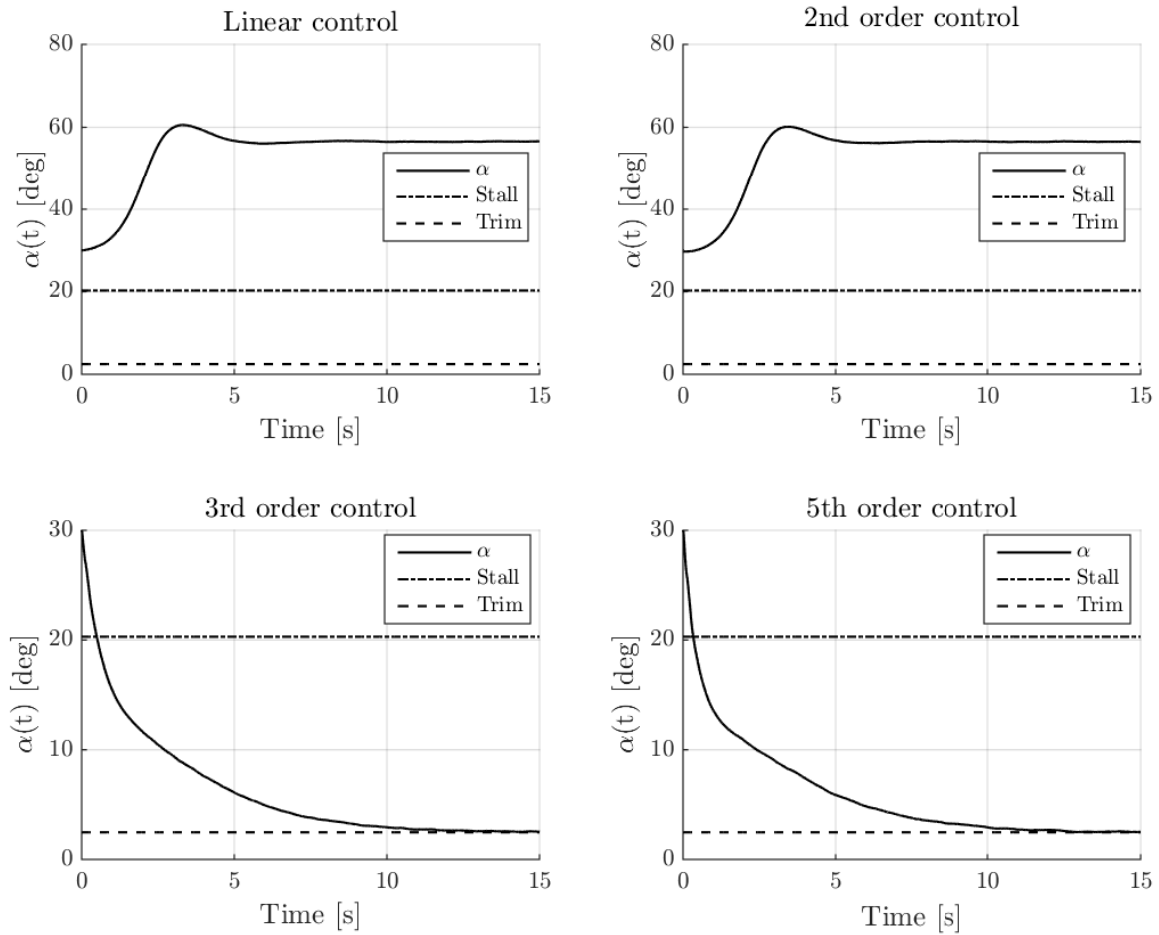}
\caption{Angle of attack of the aircraft with different control strategies. Initial condition $\alpha = 30$ deg}
\label{fig18} 
\end{figure}

\begin{figure}[h!]
\centering
\includegraphics[width=1.1\linewidth]{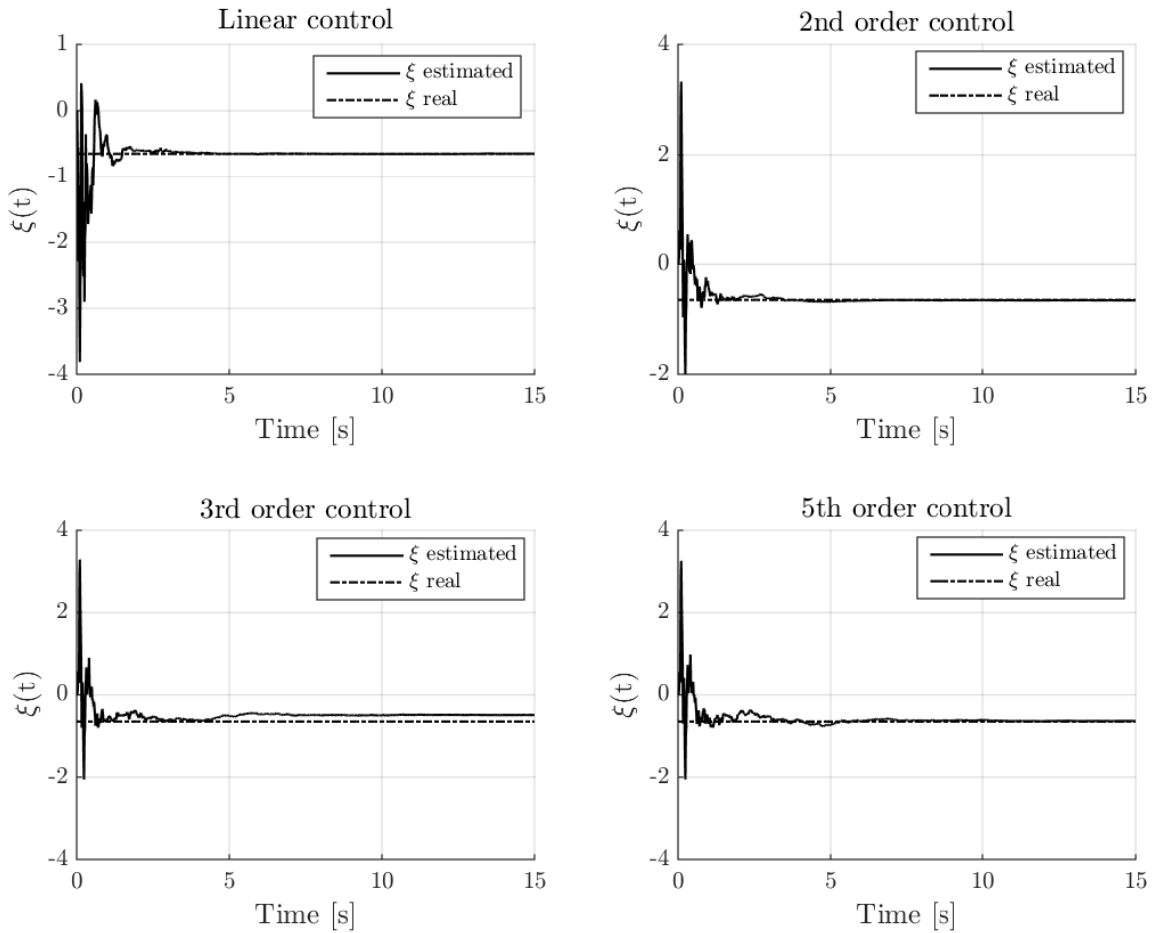}
\caption{Parameter estimation for the aircraft with different control strategies. Initial condition $\alpha = 30$ deg}
\label{fig22} 
\end{figure}

\begin{figure}[h!]
\centering
\includegraphics[width=1.1\linewidth]{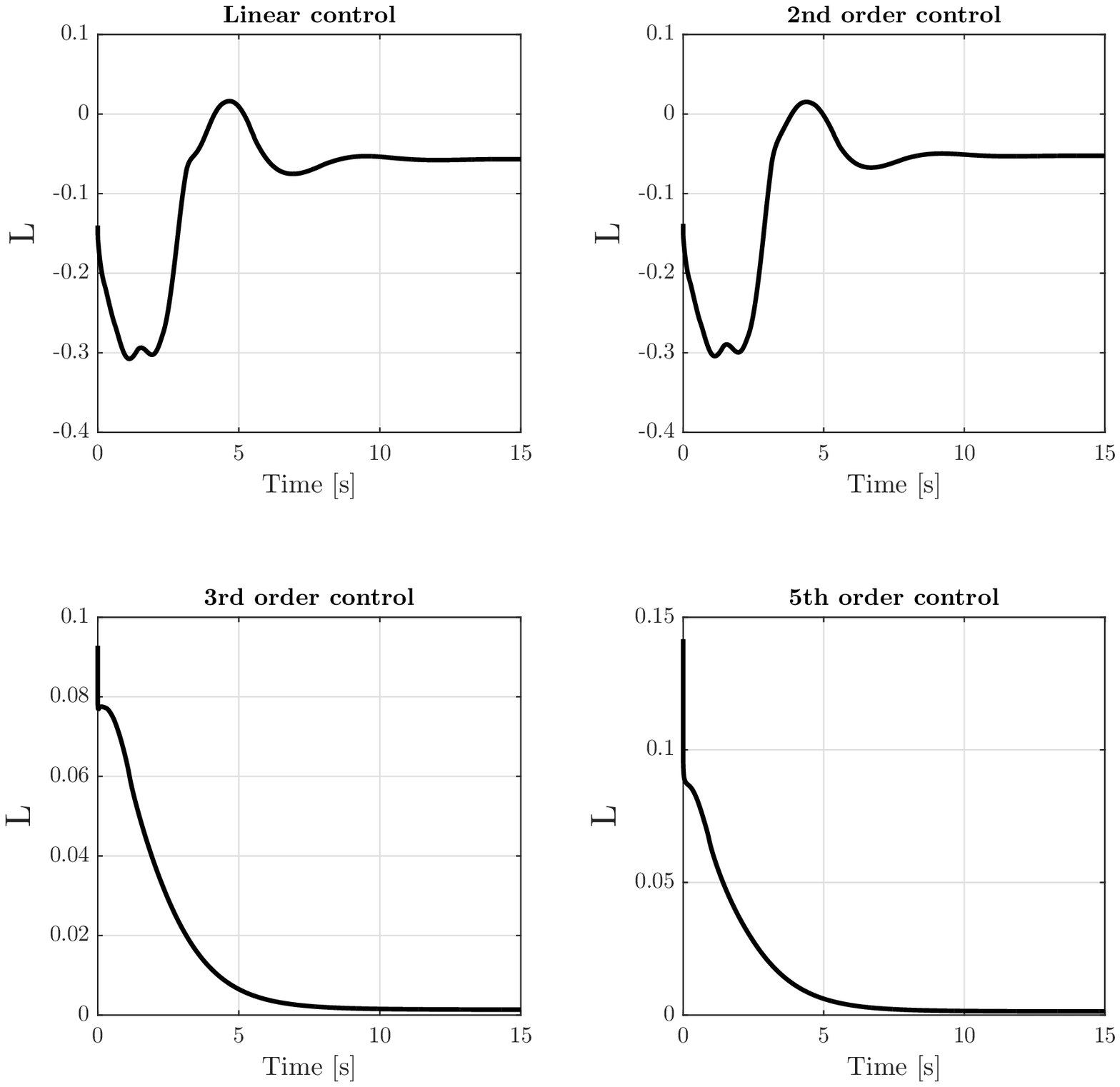}
\caption{Stability criterion verification for the aircraft with different control strategies. Initial condition $\alpha = 30$ deg}
\label{fig23} 
\end{figure}

In the second simulation, we consider $\alpha_0 = 30$ deg, which corresponds to an angle of attack within the stall region with significant loss of lift and increase of drag. Figure \ref{fig18} shows the angle of attack response for each controller. In contrast to the previous case, now the aircraft was successfully recovered from stall and reached the trim condition only for the 3rd and 5th order controllers. The 5th order controller provided a slight faster response in bringing the vehicle away from the stall condition. On the other hand, the linear and 2nd order controllers were not capable of driving the states to the trim condition. Apparently, for these control laws, the states approached another equilibrium point, but still within the stall region, at a very high angle of attack, thus not being feasible in an actual aircraft mission. 
Figure \ref{fig22} shows the parameter estimation provided by the adaptive UKF for each control strategy. We see that the estimation provided a much better approximation of $\xi$ in comparison to the previous case. The parameter estimation for the system with the linear, 2nd and 5th order controllers converged to the real value of the parameter, while the estimation for the system with the 3rd order control provided a reasonable approximation.
Figure \ref{fig23} shows the stability criterion verification for the aircraft with each controller. As expected, the $L$ function for the linear and 2nd order controls are not positive definite, therefore the trim condition is not asymptotically stable in probability in the neighborhood of the initial condition considered. The results for the system with 3rd and 5th order controllers are similar to the previous case.

Repeating the simulations for an initial angle of attack higher than $30$ deg, similar results are obtained. Therefore, we infer that, for the control design method proposed, only third or higher order controllers are capable of recovering the aircraft from very high angles of attack.
\section{Conclusions}
\label{Conclusions}
In this paper we have presented a method for studying the stability and robustness of closed loop nonlinear systems with stochastic parametric uncertainty and a strategy for designing nonlinear control laws. By using the polynomial chaos theory, we can take into account the a priori knowledge of the distribution of the random parameters  and use this information for control design purposes. The theorem presented in Section \ref{Stability} defines the conditions for the weak stochastic stability of such stochastic nonlinear systems. Furthermore, we can use the polynomial chaos expansions to propagate the uncertainty through the closed loop system and verify all possible scenarios with a low computational cost when compared with methods based on sampling. Therefore, the proposed method has clear advantages over classic methods for being less conservative and ensuring optimality, stability and robustness for nonlinear systems.

A suboptimal nonlinear feedback control was designed for controlling a high-performance aircraft with uncertain aerodynamics operating at high angles of attack. We considered a nonlinear model for the aircraft longitudinal motion as well as for the aerodynamic forces and moments. The uncertainty was introduced in the polynomial approximation of the lift coefficient curve at high angles of attack. Using the generalized polynomial chaos expansions, parameterized by random variables with known distribution, the stochastic differential equation that describes the dynamics of the system was transformed into an augmented system of deterministic differential equations for the mode strengths of the spectral decomposition. The feedback control design, formulated under the optimal control theory, was then performed in a deterministic framework. 
 
The optimal control laws were derived in a nonlinear approach. The first and second order control laws were not capable of recovering the aircraft from very high angles of attack, whereas the third and higher order controllers fulfilled the task satisfactorily. In all cases, the stability in probability of the controlled dynamics could be verified by analyzing a deterministic function. The estimation of the gPC mode strengths were performed online by the Unscented Kalman Filter in an adaptive structure.

As future work, we can work to improve the computational cost of the polynomial chaos-based controller, either in the order of expansion to have a better accuracy, or in the implementation to estimate the mode strengths in real time. In addition, the determination of the order of the expansion that will guarantee the convergence is still an open question.

\section*{Acknowledgments}
This research work is supported by the Conselho Nacional de Desenvolvimento Cient\'{i}fico e Tecnol\'{o}gico (CNPQ, Brazil), which has granted a scholarship to the first author.


\bibliographystyle{IEEEtran}

\bibliography{bibliografia}

\begin{thebibliography}{10}
\providecommand{\url}[1]{#1}
\csname url@samestyle\endcsname
\providecommand{\newblock}{\relax}
\providecommand{\bibinfo}[2]{#2}
\providecommand{\BIBentrySTDinterwordspacing}{\spaceskip=0pt\relax}
\providecommand{\BIBentryALTinterwordstretchfactor}{4}
\providecommand{\BIBentryALTinterwordspacing}{\spaceskip=\fontdimen2\font plus
\BIBentryALTinterwordstretchfactor\fontdimen3\font minus
  \fontdimen4\font\relax}
\providecommand{\BIBforeignlanguage}[2]{{%
\expandafter\ifx\csname l@#1\endcsname\relax
\typeout{** WARNING: IEEEtran.bst: No hyphenation pattern has been}%
\typeout{** loaded for the language `#1'. Using the pattern for}%
\typeout{** the default language instead.}%
\else
\language=\csname l@#1\endcsname
\fi
#2}}
\providecommand{\BIBdecl}{\relax}
\BIBdecl

\bibitem{Tol}
H.~J. Tol, C.~C. de~Visser, L.~G. Sun, E.~van Kampen, and Q.~P. Chu,
  ``Multivariate spline-based adaptive control of high-performance aircraft
  with aerodynamic uncertainties,'' \emph{Journal of Guidance and Control},
  vol.~39, no.~4, pp. 781--800, Jan. 2016.

\bibitem{Mahmood}
A.~Mahmood, Y.~Kim, and J.~Park, ``Robust $h_{\infty}$ autopilot design for
  agile missile with time-varying parameters,'' \emph{IEEE Transactions on
  Aerospace and Electronic Systems}, vol.~50, no.~4, pp. 3082 -- 3089, 2014.

\bibitem{Zollistch}
A.~W. Zollistch, F.~Holzapfel, and A.~M. Annaswamy, ``Application of adaptive
  control with closed-loop reference models to a model aircraft with actuator
  dynamics and input uncertainty,'' in \emph{American Control
  Conference}.\hskip 1em plus 0.5em minus 0.4em\relax Chicago, IL, USA: IEEE,
  Jul. 2015.

\bibitem{Zheng}
F.~Zheng and J.~Xu, ``Angle of attack control of a blended wing body aircraft
  using state feedback l1 adaptive controller,'' in \emph{Chinese Automation
  Congress}.\hskip 1em plus 0.5em minus 0.4em\relax Changsha, China: IEEE, Mar.
  2013.

\bibitem{Huang}
H.~Huang and Z.~Zhang, ``Characteristic model-based $h_2 / h_{\infty}$ robust
  adaptive control during the re-entry of hypersonic cruise vehicles,''
  \emph{Science China Information Sciences}, vol.~58, no.~1, pp. 1--21, Jan.
  2015.

\bibitem{Gavilan}
F.~Gavilan, R.~Vazquez, and J.~A. Acosta, ``Adaptive control for aircraft
  longitudinal dynamics with thrust saturation,'' \emph{Journal of Guidance,
  Control, and Dynamics}, vol.~38, no.~4, pp. 651--661, 2015.

\bibitem{Pereira}
M.~de~Freitas Virgilio~Pereira, J.~M. Balthazar, D.~A. dos Santos, A.~M.
  Tusset, D.~F. de~Castro, and I.~A.~A. Prado, ``A note on polynomial chaos
  expansions for designing a linear feedback control for nonlinear systems,''
  \emph{Nonlinear Dynamics}, vol.~87, no.~3, pp. 1653--1666, Feb. 2017.

\bibitem{Wiener}
N.~Wiener, ``The homogeneous chaos,'' \emph{American Journal of Mathematics},
  vol.~60, no.~4, pp. 897--936, 1938.

\bibitem{Xiu2002}
D.~Xiu and G.~E. Karniadakis, ``The wiener-askey polynomial chaos for
  stochastic differential equations,'' \emph{SIAM Journal on Scientific
  Computing}, vol.~24, no.~2, pp. 619--644, 2002.

\bibitem{Cameron}
R.~H. Cameron and W.~T. Martin, ``The orthogonal development of non-linear
  functionals in series of fourier-hermite functionals,'' \emph{The Annals of
  Mathematics}, vol.~48, no.~2, pp. 385--392, 1947.

\bibitem{Fisher2009}
J.~Fisher and R.~Bhattacharya, ``Linear quadratic regulation of systems with
  stochastic parameter uncertainties,'' \emph{Automatica}, vol.~45, no.~12, pp.
  2831--2841, 2009.

\bibitem{Prabhakar}
A.~Prabhakar, J.~Fisher, and R.~Bhattacharya, ``Polynomial chaos-based analysis
  of probabilistic uncertainty in hypersonic flight dynamics,'' \emph{Journal
  of Guidance and Control}, vol.~33, no.~1, pp. 222--234, 2010.

\bibitem{Xiu2003}
D.~Xiu and G.~E. Karniadakis, ``Modeling uncertainty in flow simulations via
  generalized polynomial chaos,'' \emph{Journal of Computational Physics}, vol.
  187, no.~1, pp. 137--167, 2003.

\bibitem{Abdelkefi}
A.~Abdelkefi, M.~R. Hajj, and A.~H. Nayfeh, ``Sensitivity analysis of
  piezoaeroelastic energy harvesters,'' \emph{Journal of Intelligent Material
  Systems ans Structures}, vol.~23, no.~13, pp. 1523--1531, 2012.

\bibitem{Sudret}
B.~Sudret, ``Global sensitivity analysis using polynomial chaos expansions,''
  \emph{Reliability Engineering and System Safety}, vol.~93, no.~7, pp.
  964--979, 2008.

\bibitem{Sepahvand}
K.~Sepahvand, S.~Marburg, and H.~J. Hardtke, ``Uncertainty quantification in
  stochastic systems using polynomial chaos expansions,'' \emph{International
  Journal of Applied Mechanics}, vol.~2, no.~2, pp. 305--353, 2010.

\bibitem{Ghanem}
R.~G. Ghanem and P.~D. Spanos, ``Spectral stochastic finite-element formulation
  for reliability analysis,'' \emph{Journal of Engineering Mechanics}, vol.
  117, no.~10, pp. 2351--2372, 1991.

\bibitem{Jacquelin}
E.~Jacquelin, S.~Adhikari, J.~J. Sinou, and M.~I. Friswell, ``Polynomial chaos
  expansion in structural dynamics: Accelerating the convergence of the first
  two statistical moment sequences,'' \emph{Journal of Sound and Vibration},
  vol. 356, no.~10, pp. 144--154, 2015.

\bibitem{Fisher20081}
J.~Fisher and R.~Bhattacharya, ``Stability analysis of stochastic systems using
  polynomial chaos,'' in \emph{American Control Conference (ACC)}.\hskip 1em
  plus 0.5em minus 0.4em\relax Seattle, USA: ACC, 2008.

\bibitem{Nechak}
L.~Nechak, S.~Berger, and E.~Aubry, ``Non-intrusive generalized polynomial
  chaos for the robust stability analysis of uncertain nonlinear dynamic
  friction systems,'' \emph{Journal of Sound and Vibration}, vol. 332, no.~5,
  pp. 1204--1215, 2013.

\bibitem{Kim}
K.~K. Kim and R.~D. Braatz, ``Generalized polynomial chaos expansion approaches
  to approximate stochastic receding horizon control with applications to
  probabilistic collision checking and avoidance,'' in \emph{IEEE International
  Conference on Control Applications}, IEEE, Ed., Dubrovnik, Croatia, 2012.

\bibitem{Hosder}
S.~Hosder, R.~Perez, and R.~W. Walters, ``A non-intrusive polynomial chaos
  method for uncertainty propagation in cfd simulations,'' in \emph{44th AIAA
  Aerospace Sciences Meeting and Exhibit}, AIAA, Ed., Reno, USA, 2006.

\bibitem{Ko}
J.~Ko, D.~Lucor, and P.~Sagaut, ``Effects of base flow uncertainty on couette
  flow stability,'' \emph{Computers \& Fluids}, vol.~43, no.~1, pp. 82--89,
  2011.

\bibitem{Hover}
F.~S. Hover and M.~S. Triantafyllou, ``Application of polynomial chaos in
  stability and control,'' \emph{Automatica}, vol.~42, pp. 789--795, 2006.

\bibitem{Fisher20082}
J.~Fisher and R.~Bhattacharya, ``On stochastic lqr design and polynomial
  chaos,'' in \emph{American Control Conference (ACC)}.\hskip 1em plus 0.5em
  minus 0.4em\relax Seattle, USA: ACC, 2008.

\bibitem{Peng}
Y.~B. Peng, R.~Ghanem, and J.~Li, ``Polynomial chaos expansions for optimal
  control of nonlinear random oscillators,'' \emph{Journal of Sound and
  Vibration}, vol. 329, no.~18, pp. 3660--3678, 2010.

\bibitem{Sriram}
Sriram and A.~Jameson, ``Robust optimal control using polynomial chaos and
  adjoints for systems with uncertain inputs,'' in \emph{20th AIAA
  Computational Fluid Dynamics Conference}, AIAA, Ed., Honolulu, USA, 2011.

\bibitem{PereiraASME}
M.~de~Freitas Virgilio~Pereira, I.~A.~A. Prado, D.~F. de~Castro, J.~M.
  Balthazar, R.~G.~A. da~Silva, and A.~Nabarrete, ``On nonlinear dynamics and
  flight control at high angles of attack with uncertain aerodynamics,'' in
  \emph{ASME 2016 International Mechanical Engineering Congress and
  Exposition}, ASME, Ed., Phoenix, USA, 2016.

\bibitem{Fagiano}
I.~Fagiano and M.~Khammash, ``Nonlinear model predictive control via
  regularized polynomial chaos expansions,'' in \emph{IEEE 51st Annual
  Conference on Decision and Control}, IEEE, Ed., Maui, USA, 2012.

\bibitem{Mesbah}
A.~Mesbah, S.~Streif, R.~Findeisen, and R.~Braatz, ``Stochastic nonlinear model
  predictive control with probabilistic constraints,'' in \emph{American
  Control Conference (ACC)}, IEEE, Ed., Portland, USA, 2014.

\bibitem{Templeton}
B.~A. Templeton, M.~Ahmadian, and S.~C. Southward, ``Probabilistic control
  using h2 control design and polynomial chaos: Experimental design, analysis,
  and results,'' \emph{Probabilistic Engineering Mechanics}, vol.~30, pp.
  9--19, 2012.

\bibitem{Liaw}
D.-C. Liaw and C.-C. Song, ``Analysis of longitudinal flight dynamics: a
  bifurcation-theoretic approach,'' \emph{Journal of Guidance, Control, and
  Dynamics}, vol.~24, no.~1, pp. 109--116, 2001.

\bibitem{Hui}
W.~H. Hui and M.~Tobak, ``Bifurcation analysis of aircraft pitching motions
  about large mean angles of attack,'' \emph{Journal of Guidance, Control, and
  Dynamics}, vol.~7, no.~1, pp. 113--122, 1984.

\bibitem{Jahnke}
C.~Jahnke and F.~E.~C. Culick, ``Application of bifurcation theory to the
  high-angle-of-attack dynamics of the f-14,'' \emph{Journal of Aircraft},
  vol.~31, no.~1, pp. 26--34, 1994.

\bibitem{Rom}
J.~Rom, \emph{High Angle of Attack Aerodynamics Subsonic, Transonic, and
  Supersonic Flows}.\hskip 1em plus 0.5em minus 0.4em\relax Springer, 1992.

\bibitem{Garrard2}
W.~L. Garrard, D.~F. Enns, and S.~A. Snell, ``Nonlinear feedback control of
  highly manoeuvrable aircraft,'' \emph{International journal of control},
  vol.~56, no.~4, pp. 799--812, 1992.

\bibitem{LeeAbed}
H.-C. Lee and E.~H. Abed, ``Washout filters in the bifurcation control of high
  alpha flight dynamics,'' in \emph{American Control Conference (ACC)}.\hskip
  1em plus 0.5em minus 0.4em\relax IEEE, 1991, pp. 206--211.

\bibitem{Bugajski}
D.~J. Bugajski and D.~F. Enns, ``Nonlinear control law with application to high
  angle-of-attack flight,'' \emph{Journal of Guidance, Control, and Dynamics},
  vol.~15, no.~3, pp. 761--767, 1992.

\bibitem{Wang}
Q.~Wang and R.~F. Stengel, ``Robust nonlinear flight control of a
  high-performance aircraft,'' \emph{IEEE Transactions on Control Systems
  Technology}, vol.~13, no.~1, pp. 15--26, 2005.

\bibitem{Calise}
A.~J. Calise and R.~T. Rysdyk, ``Nonlinear adaptive flight control using neural
  networks,'' \emph{Control Systems, IEEE}, vol.~18, no.~6, pp. 14--25, 1998.

\bibitem{Calafiore}
G.~C. Calafiore and F.~Dabbene, ``Probabilistic robust control,'' in
  \emph{American Control Conference (ACC)}, IEEE, Ed., vol. IEEE, New York,
  USA, 2007.

\bibitem{Branicki}
M.~Branicki and A.~J. Majda, ``Fundamental limitations of polynomial chaos for
  uncertainty quantification on systems with intermittent instabilities,''
  \emph{Communications in Mathematical Sciences}, vol. 11(1), pp. 55--103,
  2012.

\bibitem{Askey}
R.~Askey and J.~Wilson, ``Some basic hypergeometric orthogonal polynomials that
  generalize jacobi polynomials,'' \emph{Memoirs of the American Mathematical
  Society}, vol. 319, no. 319, Mar. 1985.

\bibitem{Witteveen}
J.~A.~S. Witteveen and H.~Bijl, ``Modeling arbitrary uncertainties using
  gram-schmidt polynomial chaos,'' in \emph{44th AIAA Aerospace Sciences
  Meeting and Exibit}, 2006.

\bibitem{Papoulis}
A.~Papoulis and S.~U. Pillai, \emph{Probability, Random Variables, and
  Stochastic Processes}, 4th~ed.\hskip 1em plus 0.5em minus 0.4em\relax
  McGraw-Hill Higher Education, 2002.

\bibitem{Debusschere}
B.~J. Debusschere, H.~N. Najm, P.~P. Pebay, O.~M. Knio, R.~G. Ghanem, and
  O.~P.~L. Maitre, ``Numerical challenges in the use of polynomial chaos
  representations for stochastic processes,'' \emph{SIAM Journal on Scientific
  Computing}, vol.~26, no.~2, p. 698 719, 2004.

\bibitem{Khasminskii}
R.~Khasminskii, \emph{Stochastic Stability of Differential Equations},
  2nd~ed.\hskip 1em plus 0.5em minus 0.4em\relax Springer, 2010.

\bibitem{ChenChen}
G.~Chen, G.~Chen, and S.~H. Hsu, \emph{Linear Stochastic Control
  Systems}.\hskip 1em plus 0.5em minus 0.4em\relax CRC Press, 1995.

\bibitem{Kushner}
H.~J. Kushner, \emph{Stochastic Stability and Control}, R.~Bellman, Ed.\hskip
  1em plus 0.5em minus 0.4em\relax Academic Press Inc., 1967.

\bibitem{Bryson}
A.~E. Bryson and Y.~C. Ho, \emph{Applied Optimal Control: Optimization,
  Estimation and Control}.\hskip 1em plus 0.5em minus 0.4em\relax Taylor \&
  Francis Group, 1975.

\bibitem{Garrard1967}
W.~L. Garrard, N.~H. McClamroch, and L.~G. Clark, ``An approach to sub-optimal
  feedback control of non-linear systems,'' \emph{International Journal of
  Control}, vol.~5, no.~5, pp. 425--435, 1967.

\bibitem{Garrard}
W.~L. Garrard and J.~M. Jordan, ``Design of nonlinear automatic flight control
  systems,'' \emph{Automatica}, vol.~13, no.~5, pp. 497--505, 1977.

\bibitem{Garrard1969}
W.~L. Garrard, ``Additional results on sub-optmal feedback control of
  non-linear systems,'' \emph{International Journal of Control}, 1969.

\bibitem{Lukes}
D.~L. Lukes, ``Optimal regulation of nonlinear dynamical systems,'' \emph{SIAM
  Journal on Control}, vol.~7, no.~1, pp. 75--100, Feb. 1969.

\bibitem{Markus}
E.~B. Lee and L.~Markus, \emph{Foundations of Optimal Control Theory}.\hskip
  1em plus 0.5em minus 0.4em\relax New York: John Wiley, 1967.

\bibitem{Ljung}
L.~Ljung and T.~Soderstrom, \emph{Theory and Practice of Recursive
  Identification}.\hskip 1em plus 0.5em minus 0.4em\relax MIT Press, 1983.

\bibitem{Goodwin}
G.~C. Goodwin and K.~S. Sin, \emph{Adaptive Filtering Prediction and
  Control}.\hskip 1em plus 0.5em minus 0.4em\relax Dover Publications Inc,
  1984.

\bibitem{Narendra}
K.~S. Narendra and A.~M. Annaswamy, \emph{Stable Adaptive Systems}.\hskip 1em
  plus 0.5em minus 0.4em\relax Dover Publications Inc., 2005.

\bibitem{Dutta}
P.~Dutta and R.~Bhattacharya, ``Nonlinear estimation with polynomial chaos and
  higher order moment updates,'' in \emph{American Control Conference}.\hskip
  1em plus 0.5em minus 0.4em\relax Baltimore, USA: IEEE, Jun. 2010.

\bibitem{Liu}
J.~Liu, Y.~Wang, and J.~Zhang, ``A linear extension of unscented kalman filter
  to higher-order moment-matching,'' in \emph{IEEE 53rd Annual Conference on
  Decision and Control (CDC)}.\hskip 1em plus 0.5em minus 0.4em\relax Los
  Angeles, USA: IEEE, Feb. 2014.

\bibitem{Ponomareva}
K.~Ponomareva, P.~Date, and Z.~Wang, ``A new unscented kalman filter with
  higher order moment-matching,'' in \emph{19th International Symposium on
  Mathematical Theory of Networks and Systems}.\hskip 1em plus 0.5em minus
  0.4em\relax Budapest, Hungary, Jul. 2010.

\bibitem{Grothe}
O.~Grothe, ``A higher order correlation unscented kalman filter,'' \emph{A
  higher order correlation unscented Kalman filter}, vol. 219, no.~17, pp.
  9033--9042, 2013.

\bibitem{Majji}
M.~Majji, J.~L. Junkins, and J.~D. Turner, ``A high order method for estiamtion
  of dynamic systems,'' \emph{The Journal of the Astronautical Sciences},
  vol.~56, no.~3, pp. 401--440, 2008.

\bibitem{Atesoglu}
O.~Atesoglu, ``High angle of attack maneuvering and stabilization control of
  aircraft,'' Ph.D. dissertation, Middle East Technical University, 2007.

\bibitem{Anderson}
J.~D. Anderson, \emph{Fundamentals of Aerodynamics}, 5th~ed.\hskip 1em plus
  0.5em minus 0.4em\relax McGraw-Hill Education, 2010.

\bibitem{PereiraBaltha}
D.~C. Pereira, J.~M. Balthazar, F.~R. Chavarette, and M.~Rafikov, ``On
  nonlinear dynamics and an optimal control design to a longitudinal flight,''
  \emph{Journal of Computational and Nonlinear Dynamics}, vol.~3, no.~1, p.
  011012, 2008.

\bibitem{Lewis}
B.~L. Stevens and F.~L. Lewis, \emph{Aircraft Control and Simulation},
  2nd~ed.\hskip 1em plus 0.5em minus 0.4em\relax John Wiley \& Sons, Inc.,
  2003.

\end{thebibliography}

\begin{IEEEbiography}[{\includegraphics[width=1in,height=1.25in,clip,keepaspectratio]{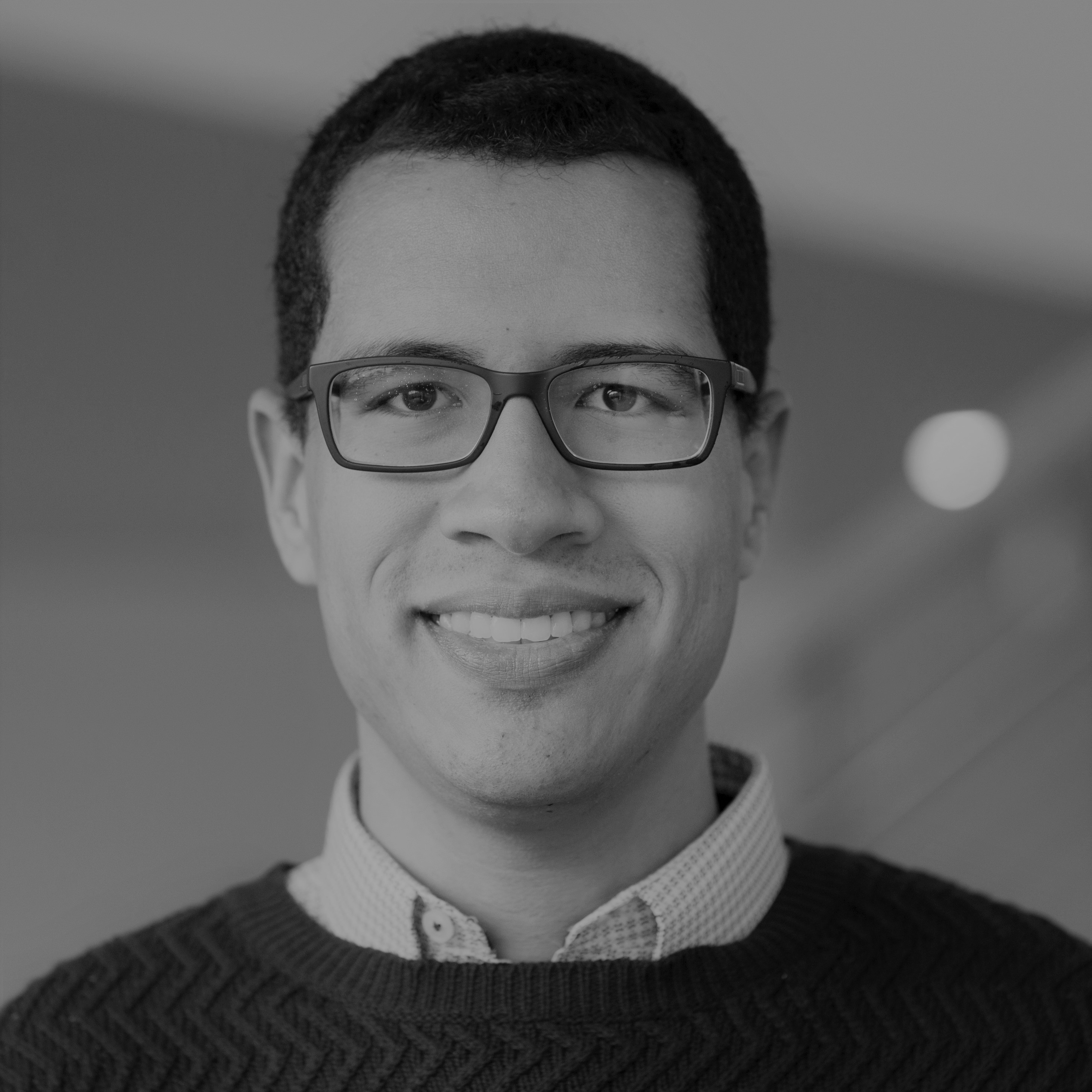}}]{Matues de F. V. Pereira}
received his B.S. degree in Aerospace Engineering from the Federal University of Minas Gerais - UFMG, Brazil, in 2015 and his M.Sc. degree in Aeronautical and Mechanical Engineering from the Aeronautics Institute of Technology - ITA, Brazil, in 2017. He is currently a PhD candidate in Aerospace Engineering at the University of Michigan, USA.
His research interests include optimal control theory, stochastic uncertainty propagation, nonlinear dynamics and stability with applications in aerospace and mechanical systems.
\end{IEEEbiography}

\begin{IEEEbiography}[{\includegraphics[width=1in,height=1.25in,clip,keepaspectratio]{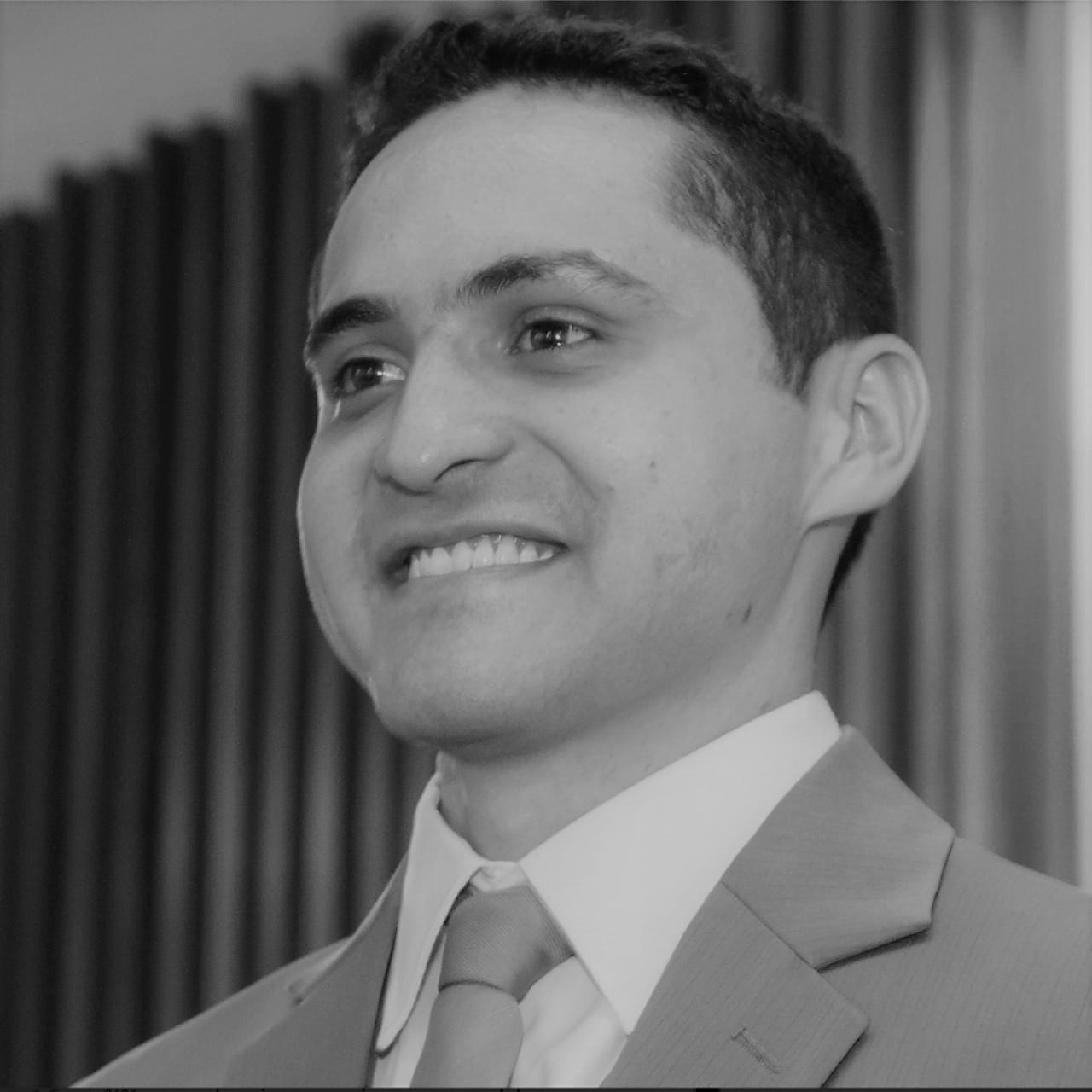}}]{Igor A. A. Prado}
received his B.S. degree in Control and Automation Engineering from the University of the State of Amazonas - UEA, Manaus - AM, Brazil, in 2011 and his M.Sc. degree in Aeronautical and Mechanical Engineering from the Aeronautics Institute of Technology - ITA, S\~{a}o Jos\'{e} dos Campos-SP, Brazil, in 2014, where he is currently pursuing his Ph.D. degree in the Aeronautical and Mechanical Engineering Department.  His current research interests include optimal control, predictive control, discrete mechanics and robotic systems.
\end{IEEEbiography}


\begin{IEEEbiography}[{\includegraphics[width=1in,height=1.25in,clip,keepaspectratio]{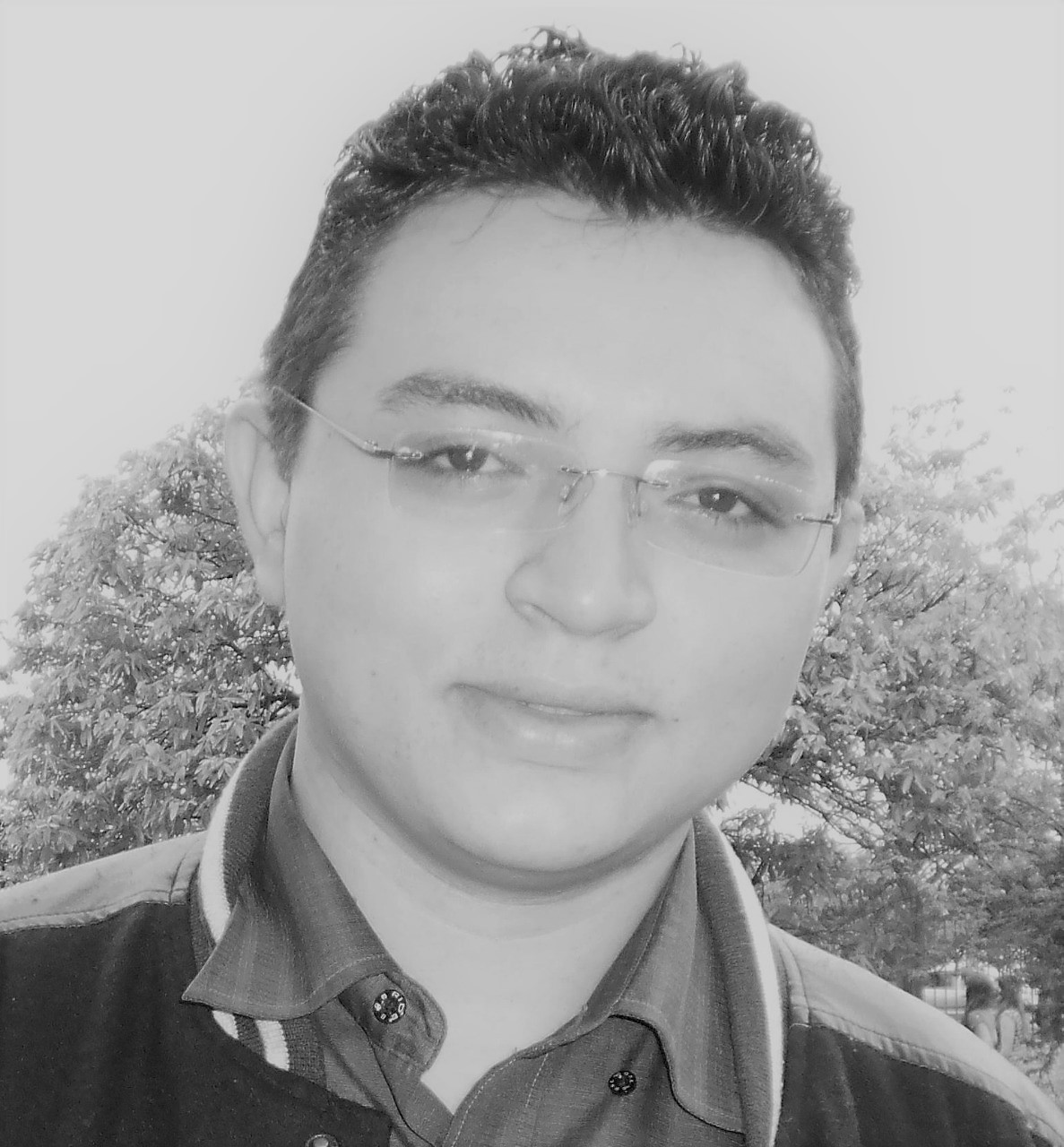}}]{Davi F. de Castro} received his B.S. degree in Control and Automation Engineering from the University of the State of Amazonas - UEA, Manaus - AM, Brazil, in 2011 and his M.Sc. degree in Aeronautical and Mechanical Engineering from the Aeronautics Institute of Technology - ITA, S\~{a}o Jos\'{e} dos Campos-SP, Brazil, in 2014, where he is currently pursuing his Ph.D. degree in the Aeronautical and Mechanical Engineering Department. 
His current research interests include control, modeling, identification, discrete mechanics and robotic systems.
\end{IEEEbiography}

\begin{IEEEbiography}[{\includegraphics[width=1in,height=1.25in,clip,keepaspectratio]{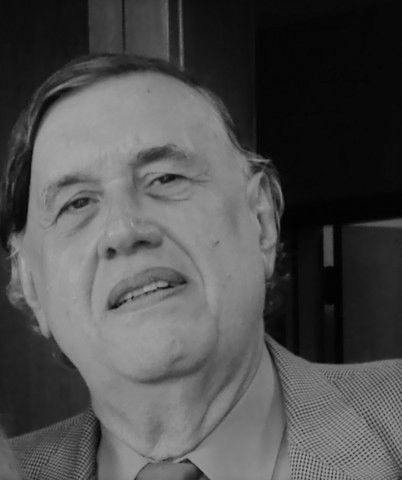}}]{Jos\'{e} M. Balthazar}
received his B.S. degree in Applied Mathematics Universidade Estadual Paulista - UNESP, Brazil in 1971, his M.Sc. degree in Sciences from the Aeronautics Institute of Technology - ITA, Brazil, in 1975, and his PhD degree in Mechanical Engineering from the University of S\~{a}o Paulo - USP, Brazil, in 1993. He is currently a Professor at the Federal University of Technology Paran\'{a}, Ponta Grossa, Brazil.
His research interests include nonlinear dynamics, chaos and control applications in engineering.
\end{IEEEbiography}




\end{document}